\DeclareUnicodeCharacter{2212}{-}
\documentclass[journal=jctcce,manuscript=article]{achemso}
\setkeys{acs}{usetitle = true}
\setkeys{acs}{maxauthors=10,etalmode=truncate}
\usepackage{xspace, amsmath, multirow, setspace, graphicx, subfigure}
\usepackage{bm, dcolumn, epsfig, color, xr, booktabs, longtable, lscape, titlecaps}
\Addlcwords{the of into to in for a an with on and at from by}
\mciteErrorOnUnknownfalse
\title[]{Solvent Effects on the Menshutkin Reaction}

\author{Haydar Taylan Turan} \affiliation{Department of Chemistry,
  University of Basel, Klingelbergstrasse 80, Basel, Switzerland}

\author{Sebastian Brickel} \affiliation{Department of Chemistry,
  University of Basel, Klingelbergstrasse 80, Basel, Switzerland}
\affiliation{Present Address: Department of Chemistry - BMC, Uppsala
  University, BMC Box 576, 751 23 Uppsala, Sweden}

\author{Markus Meuwly} \email{m.meuwly@unibas.ch}
\affiliation{Department of Chemistry, University of Basel,
  Klingelbergstrasse 80, Basel, Switzerland}


\begin{document}
\doublespace

\maketitle
\thispagestyle{empty}

\maketitle

\begin{abstract}
The Menshutkin reaction is a methyl transfer reaction relevant in
fields ranging from biochemistry to chemical synthesis. In the present
work, energetics and solvent distributions for NH$_{3}$+MeCl and
Pyr+MeBr reactions were investigated in the gas-phase, in water,
methanol, acetonitrile, benzene, and in cyclohexane by means of
reactive molecular dynamics simulations. For polar solvents (water,
methanol, and acetonitrile) and benzene, strong to moderate catalytic
effect for both reactions is found whereas apolar and bulky
cyclohexane interacts weakly with the solute and does not show
pronounced barrier reduction. Calculated barrier heights for the
Pyr+MeBr reaction in acetonitrile and cyclohexane are 23.2 and 28.1
kcal/mol compared with experimentally measured barriers of 22.5 and
27.6 kcal/mol, respectively. The 2-dimensional solvent distributions
change considerably between reactant and TS but comparatively little
between TS and product conformations of the solute. The simulations
also suggest that as the system approaches the transition state,
correlated solvent motions that destabilize the solvent-solvent
interactions are required to surmount the barrier. Finally, the
average solvent-solvent interaction energies in the reactant, TS, and
product state geometries are correlated with changes in the solvent
structure around the solute.
\end{abstract}

\section{Introduction}
Solvation is essential in chemistry and directly influences properties
such as reaction rates or spectroscopic responses of
solutes. \cite{Warshel1980,Yadav1991,Luzhkov1991,Zhan2017,Kamerlin2009,Ranaghan2004,Shaw2010,Carlsson2006,Bingemann2000,Almlof2007,Elles2004,Preston2013}
Depending on the nature of the solvent (e.g. polar or apolar) the
potential energy surface underlying the nuclear dynamics changes and
affects mechanistic aspects of the
dynamics.\cite{Rivera2011,Orr-Ewing2014,Claeyssens2011} As an example,
for the Claisen rearrangement the reaction barrier in polar solvent
decreases compared to the gas
phase\cite{severance.jacs.1992.ave,guest.perk2.1997.ave,Cramer1992,MM.claisen:2019}
whereas it increases for S$_{\rm N}$2 reactions in going from the gas
phase into
solution.\cite{Hwang1988,Shaik1985,Chandrasekhar1985,Merkel1988,Gao1993,Fradera1996,Adamovic2005,Shaik2006}\\

\begin{figure}[H]
\includegraphics[width=0.75\textwidth]{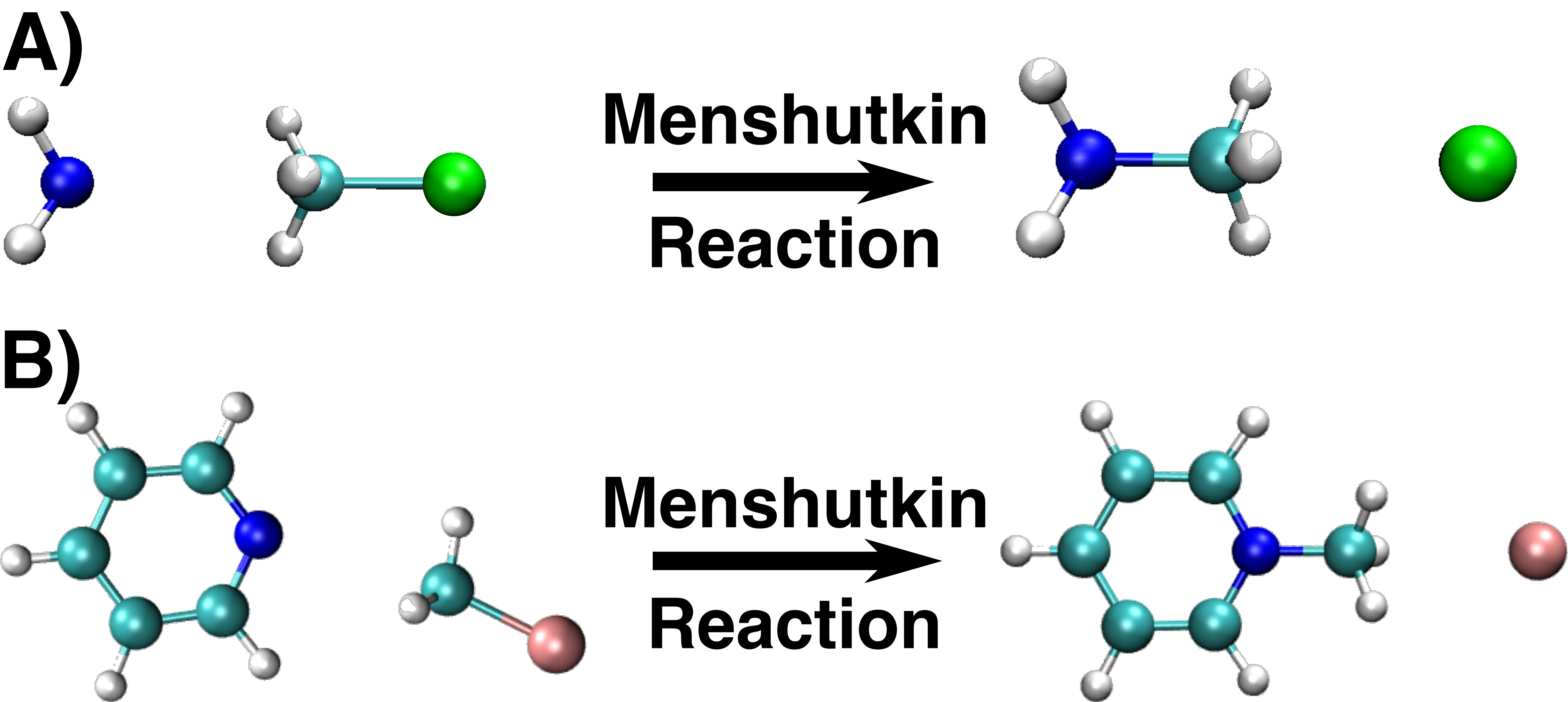}
\caption{Schematic representation for A) the Menshutkin reaction of
  ammonia with chloromethane and B) the Menshutkin reaction of
  pyridine with bromomethane.}
\label{fig:fig1}
\end{figure}

\noindent
The Menshutkin reaction\cite{Menschutkin1890a,Menschutkin1890} is an
important methyl transfer reaction and plays an essential role in many
fields of chemistry ranging from biochemical processes (histone
methylation\cite{allfrey:1964},
methyltransferases\cite{Schubert2003,Schmidt2014}) to chemical
synthesis.\cite{arava2017strategies} Methyl-transfer reactions in
biological systems are catalyzed by methyltransferases for which the
most prominent enzymatic cofactor is $S$-adenosyl-L-methionine
(SAM).\cite{Schubert2003} SAM is derived from ATP by the enzyme
methionine synthase\cite{ding2015s} and serves as a regulator of a
variety of processes including DNA, tRNA, and rRNA methylation as well
as immune response.\cite{scavetta2000structure} DNA methylation is
governed by DNA methyltransferase (DNMTs) enzymes such as DNMT1,
DNMT3A, and DNMT3B.\cite{tajima2016domain} More recent findings
suggest that DNA and histone methylation cooperate to maintain the
cellular epigenomic landscape.\cite{li2021interplay} Finally, histone
methylation has been found to be undone by histone
demethylases.\cite{shi:2004}\\

\noindent
Common examples of methylation reactions that have been studied
extensively in the past are S$_N$2
reactions\cite{Lieder1975,Olmstead1977,Hierl1987,Merkel1988,Barlow1988,VandeLinde1989,DePuy1990,Kato2001,Stei2016,Carrascosa2017}
such as Cl$^{-}$ + MeBr $\rightarrow$ MeCl + Br$^{-}$
\cite{Adamovic2005} or F$^{-}$ + MeI $\rightarrow$ MeF + I$^{-}$
\cite{Stei2016} and the Menshutkin
reaction.\cite{Viers1984,Amovilli1998,Castejon2001,Fradera1996,Sola1991,Jiang2013,Acevedo2010,Castejon1999}
The S$_N$2 and Menshutkin reactions are of particular interest because
of their importance in chemical synthesis.\cite{Gronet1991} The main
difference between a standard S$_N$2 reaction and a Menshutkin
reaction is that for an S$_N$2 reaction the reactant state is charged
whereas it is neutral for a Menshutkin reaction, e.g. NH$_{3}$ +
NH$_{2}$−C$_{6}$H$_{4}$-Br $\rightarrow$
NH$_{2}$-C$_{6}$H$_{4}$-NH$_{3}^{+}$+Br$^{-}$ or Pyr+MeCl
$\rightarrow$
[Pyr-Me]$^{+}$+Cl$^{-}$.\cite{Jiang2013,Castejon1999}. The Menshutkin
reaction proceeds via a dipolar transition state (TS) to an ionic
product and the TS is stabilized in polar
solvents.\cite{Castejon2001,Castejon1999}\\

\noindent
The Menshutkin reaction is also a characteristic example for studying
the effect of solvation on reactions\cite{Acevedo2010} since it
displays pronounced solvent effects on activation
barriers.\cite{Castejon1999, auriel1979quantitative,
  maran1994semiempirical} Experimental studies have shown a decrease
in the activation energy depending on the polarity of the solvent. The
experimentally reported activation barrier for Pyr+MeBr in methanol is
22.5 kcal/mol which is 3.1 and 5.1 kcal/mol lower compared with apolar
cyclohexane and di-n-butyl ether\cite{Castejon1999}. Further, an
almost linear relationship between the natural log of the rate
constants and polarity/polarizability of the solvents for the
alkylation (Menshutkin) reaction between 1-bromodecane and
1,2-dimethylimidazol was observed.\cite{komarova2021diformylxylose}
DMSO was found to be the best solvent for accelerating the rate of
heteroatom alkylations. Although solvent effects are evidently
relevant for the Menshutkin reaction, surprisingly little is known
about the molecular details of the reaction itself and the solvent
ordering in particular. \\

\noindent
Solvent effects for the NH$_{3}$+MeCl and Pyr+MeBr reactions were
previously considered with quantum mechanical (QM) and semi-empirical
methods. For NH$_{3}$+MeCl previous electronic structure calculations
at the HF/6-31++G*\cite{poater2001effects} and B3PW91/6-31+G*
\cite{halls2002chemistry} levels with implicit polarizable continuum
model (PCM) models\cite{miertuvs1981electrostatic} as well as QM/MM
simulations at the B3LYP/MM level were carried
out.\cite{pan2019accelerated} Also, both reactions were investigated
with multiscale reaction density functional theory (RxDFT) in
acetonitrile.\cite{tang2020solvent} \\

\noindent
The present work addresses the question of how the presence and the
nature of the solvent affects the reaction barrier height for two
methyl transfer reactions NH$_{3}$+MeCl and Pyr+MeBr (see
Figure~\ref{fig:fig1}). For a comprehensive sampling of the molecular
dynamics along the progression coordinate the multi state adiabatic
reactive molecular dynamics (MS-ARMD) framework was used which
provides a computational means to investigate chemical reactions (bond
breaking/bond formation) with an efficiency comparable to an empirical
force field.\cite{nagy.jctc.2014.msarmd} Both, the energetics of the
reaction and the organization of the solvent for critical points along
the progression coordinate is analyzed. First, the force fields and
their parametrization is described. This is followed by free energy
calculations and the analysis of the catalytic effect, the
organization of the solvent around the transition state and the
solvent energetics. Finally, the results are discussed in a broader
context.\\

\section{Computational Methods}
\subsection{Molecular Dynamics Simulations}
All molecular dynamics (MD) simulations were performed with
CHARMM\cite{charmm.prog} with provision for forming and breaking bonds
via MS-ARMD.\cite{nagy.jctc.2014.msarmd} MD simulations were started
following 500 steps of steepest descent and 500 steps of Adopted Basis
Newton-Raphson minimization. Then, 500 ps of $NpT$ dynamics were
carried out using the leapfrog Verlet integrator
\cite{verlet1967computer} ($\Delta t = 1$ fs) and a Hoover
barostat\cite{hoover1985canonical} with a collision rate of 5
ps$^{-1}$. The $NpT$ dynamics were followed by 2 ns ($\Delta t = 1$
fs) of free dynamics in the $NVT$ ensemble using
SHAKE\cite{W_VanGunsteren_MolPhys_1977}. Periodic Boundary Conditions
(PBC) together with the Particle Mesh Ewald (PME\cite{Essmann1995})
method were used for the long range electrostatic interaction. The
cut-off for non-bonded, switching, and smoothing was 16 \AA, 14 \AA,
and 12 \AA, respectively.\\

\noindent
The simulations in the different solvents were started by solvating
the reactant states of the two reactions, NH$_3$+MeCl and Pyr+MeBr, in
pre-equilibrated, cubic solvent boxes. Due to the different shapes and
sizes of the solvents, cubic box sizes were chosen accordingly. They
were of length $L=30$ \AA\/ (water), $L=25$ \AA\/ (methanol),
$L=28$ \AA\/ (acetonitrile), $L=27$ \AA\/ (benzene), and $L=30$ \AA\/
(cyclohexane), respectively.\\

\subsection{Parametrization of the Force Field}
All electronic structure calculations of reactant, transition, and
product states were performed with Gaussian09\cite{g09} at the
MP2/6-311++G(2d,2p) level of theory. For simulating Me-transfer
between a donor and an acceptor a reactive force field to break and
form chemical bonds is required. MS-ARMD is a reactive molecular
dynamics implementation which combines individually weighted
connectivities, describing different states by parametrized force
fields. So called, GAPOs (GAussian $\times$ POlynomials) for
describing the adiabatic barrier are added to smoothly connect
reactant and product side to form the global reactive Potential Energy
Surface (PES). GAPOs are calculated from the energy difference $\Delta
V_{ij}(x)$ ($=V_j(x)-V_i(x)$) between two states $i$ and $j$.\\

\noindent
For the Menshutkin reactions NH$_{3}$+MeCl and Pyr+MeBr the
parametrization of the reactant and product states started with
optimization of the reactant and product geometries at the
MP2/6-311++G(2d,2p) level of theory. Initial force fields for the
reactant and product molecules (NH$_{3}$, MeCl, methylammonium cation,
pyridine, and methyl pyridine cation, respectively), were obtained
from Swissparam.\cite{zoete.jcc.2011.swissparam} The initial Mulliken
charges for the fit were those from the electronic structure
calculations.\\

\noindent
For NH$_3$+MeCl the reference reactant complex structures were
generated from a 1 ns gas-phase MD simulation in the $NVT$ ensemble
using the Swissparam parametrization. In the MD simulations the
distance between the ammonia and chloromethane was kept close to the
equilibrium distance in the reactant state using a harmonic
constraint. From this trajectory 1000 structures were stored for which
reference energies at the MP2/6-311++G(2d,2p) level were
determined. For the Pyr+MeBr reaction a similar procedure was
followed. The reference reactant complex structures were generated
from a 1 ns gas-phase MD simulation in the $NVT$ ensemble using the
Swissparam parametrization. The distance between the pyridine and
bromomethane was kept close to the equilibrium distance in the
reactant state using a harmonic constraint. From this trajectory 1700
structures were stored for which reference energies at the
MP2/6-311++G(2d,2p) level were determined.\\

\noindent
In addition to the conformational ensemble of the reactant states,
structures and energies along the internal reaction coordinate (IRC)
path were used together with product state geometries which were
generated from scanning along the C--Cl and C--Br distances,
respectively. As the main focus in the present work is on the
energetics involving the reactant, TS, and ion-pair states, no full
parametrization for the ionic products [H$_3$NMe]$^+$ and
[PyrMe]$^{+}$ was carried out. This would be required if the reaction
is followed out and beyond the solvent-separated ion pairs which is
not done in the following. Hence, on the product side the relaxed scan
only involved C--Cl separations between 2.5 \AA\/ to 6 \AA\/ and C--Br
distances ranging from 3 \AA\/ to 13 \AA\/, respectively. Subsequently
all force field parameters were fitted to the reference
energies\cite{Brickel2019,MM.claisen:2019,Brickel2017,YosaReyes2016}
by using a downhill simplex algorithm\cite{Nelder1965}. Finally, the
reactant and product force fields were connected by fitting the GAPOs
along the IRC using a genetic algorithm.  The bond, angle, dihedral
and van der Waals parameters of the force field are provided in
Tables S1 and S2. Further, the resulting fit
of the Menshutkin reactions for NH$_{3}$+MeCl and Pyr+MeBr contains
three Gaussians and the GAPO parameters are provided in Tables
S3 and S4.\\

\subsection{Umbrella sampling}
Since direct sampling of the two reactions is not possible due to the
high reaction barriers ($> 20$ kcal/mol), umbrella sampling
(US)\cite{kottalam.jacs.1988.us} simulations were used to follow the
reaction path. The reaction coordinate chosen here was the difference
between the carbon-X bond of the reactant and the carbon-N bond in the
product, i.e. $r_c = d_{\text{C}\text{X}} - d_{\text{C}\text{N}}$
where X = Cl, Br for the NH$_{3}$+MeCl and Pyr+MeBr reactions,
respectively. Simulations were carried out for equidistant windows
between $r_c = -1.3$ and 1.6 \AA\/ with $\Delta r_c = 0.1$ \AA\/ and
with $k_{\text{umb}} = 150$ kcal/mol/rad$^2$.\\

\noindent
US simulations for both systems were performed in a sequential manner,
i.e. restarting a new US simulation from the structure and velocities
of the previous simulation. Each window was simulated for 50 ps and
window statistics was accumulated after equilibration for 5 ps.
Statistics from all the windows were combined to yield the
1-dimensional potential of mean force (PMF) using the Weighted
Histogram Analysis Method
(WHAM)\cite{Kumar1992,souaille.comphycom.2001.us} with a tolerance of
0.001.\\

\subsection{Solvent Distribution}
To characterize the solvent distribution and change along the reaction
pathway, separate 2 ns MD simulations for NH$_{3}$+MeCl and Pyr+MeBr
in the \textit{NVT} ensemble were carried out. The reactant, TS and
product state structures were constrained at their MP2/6-311++G(2d,2p)
optimized geometries and the center of mass (CoM) of the solute was at
the center of the simulation box. To analyze the solvent distributions
the solute was reoriented and superimposed to minimize the structural
root mean squared deviation. For the NH$_{3}$+MeCl reaction first the
methyl-carbon was translated to the origin of the box, then the
ammonia-N was aligned along the $x-$axis and one of the three
hydrogens bonded to the ammonia-N was placed into the $xy-$plane. For
the Pyr+MeBr reaction the pyridine-N was translated to the origin of
the box, the methyl-C was then aligned along the $x-$axis and one of
the hydrogens bonded to the ammonia-N was placed in the $xy-$plane. \\

\noindent
From the 2000 snapshots the water-oxygen (O$_{\rm W}$),
methanol-oxygen (O$_{\rm MeOH}$) and acetonitrile-nitrogen (N$_{\rm
  ACN}$) coordinates were extracted whereas for benzene and
cyclohexane the center of mass was determined and used for further
analysis. The CoM was chosen for apolar solvents because selecting one
specific atom as for the polar solvents is less meaningful as this
would lead to skewing the results due to rotation of the solvent
molecule. Based on these reference coordinates, all solvent molecules
with the point of reference (O$_{\rm W}$, O$_{\rm MeOH}$, N$_{\rm
  ACN}$, and the CoMs of benzene and hexane) within $-1 \leq z \leq
+1$ \AA\/ were selected and further analyzed to give a 2-dimensional
solvent distribution. The collected data points were then smoothed
with a 2D bivariate normal kernel density
estimation\cite{rosenblatt1956remarks,parzen1962estimation}. Solvent
boxes were divided into 100 bins along the $x-$ and $y-$axes to
generate a 100 $\times$ 100 grid. Then, the solvent density around
each of the grid points was smoothed using a 2D Gaussian distribution
with a width of 1.5 \AA\/. Isocontours were drawn at 90\%, 75\%, 50\%,
25\%, and 10\% of the highest occupation for each solvent to allow
direct comparison.\\

\section{Results and Discussion}

\subsection{The Potential Energy Surfaces}
\begin{figure}[H]
\centering
\includegraphics[width=0.48\linewidth]{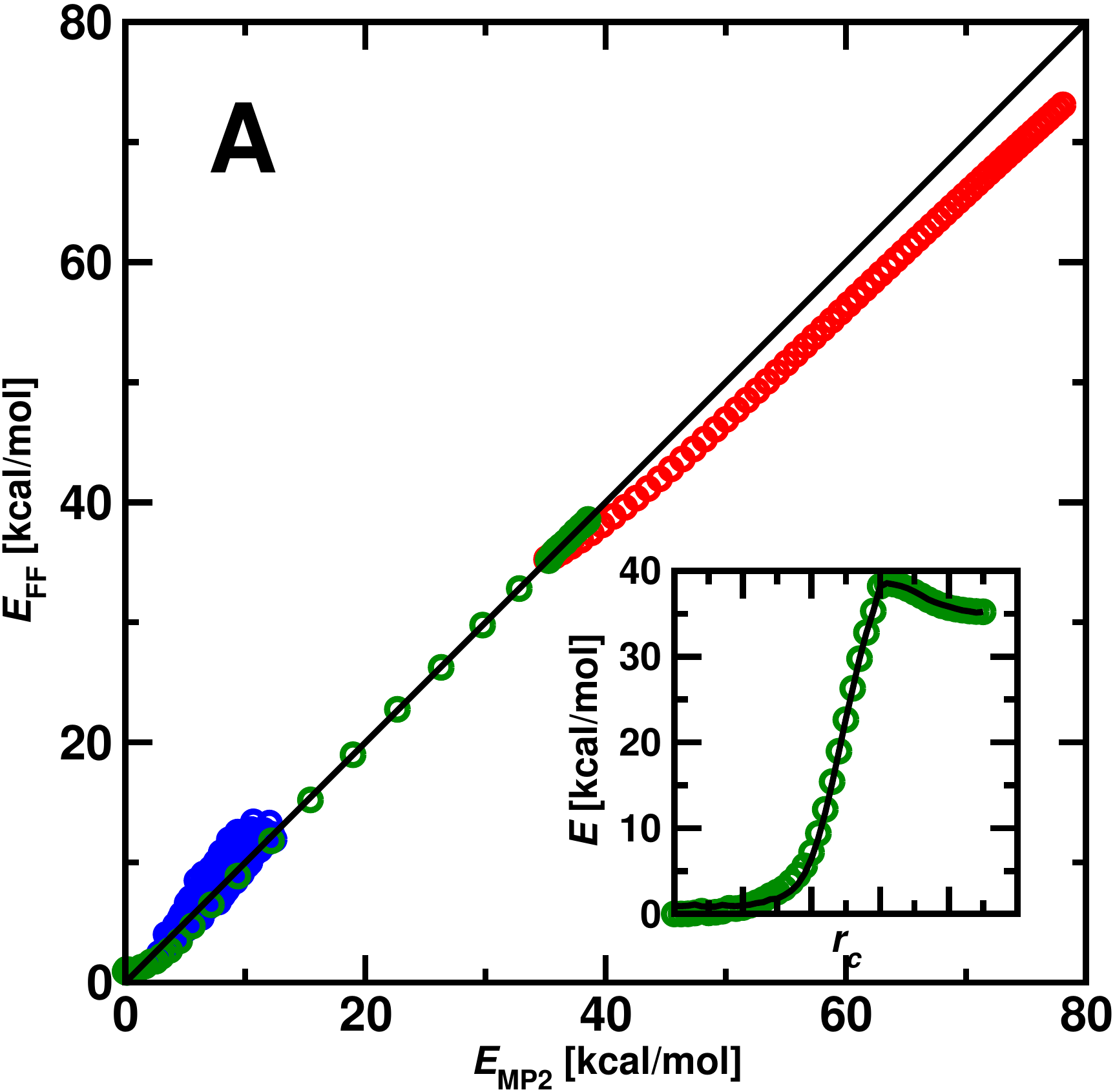}
\includegraphics[width=0.47\linewidth]{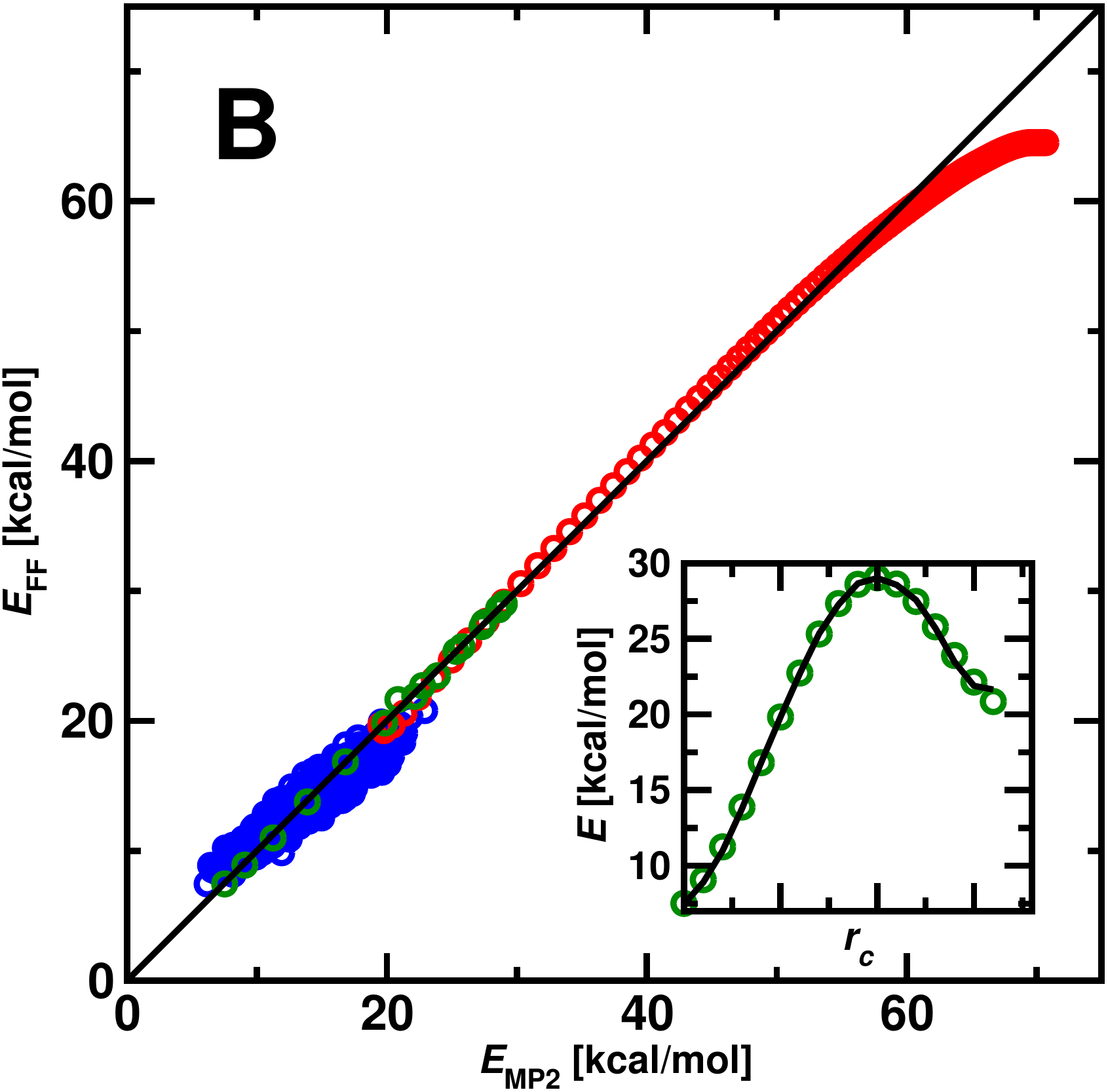}
\caption{Panel A: Menshutkin reaction of ammonia with chloromethane to
  methylammonium chloride: Energy correlation between the fitted force
  field and the MP2 reference for 1000 reactant structures (blue, RMSD
  of 0.78 kcal/mol) and 70 product structures (red, RMSD of 3.60
  kcal/mol). Panel B: Menshutkin reaction of pyridine with
  bromomethane to methyl pyridine and Br$^-$: Energy correlation
  between the fitted force field and the MP2 reference for 1700
  reactant structures (blue, RMSD=0.93 kcal/mol) and 101 product
  structures (red, RMSD=2.41 kcal/mol).  The two insets show the IRC
  points (green circles) together with the energies from the reactive
  force field (black line) based on the GAPO-fits to connect the
  reactant and product force fields.}
\label{fig:pes}
\end{figure}

The correlation between the fitted force field and the MP2 reference
energies is reported in Figure~\ref{fig:pes}. As the main interest of
the present work is the forward process, i.e. NH$_{3}$+MeCl
$\rightarrow$ [H$_3$NMe]$^+$+Cl$^{-}$ and Pyr+MeBr $\rightarrow$
[PyrMe]$^{+}$+Br$^{-}$, the parametrization was focused on the
reactant state including the transition state region. For the 1000 and
1700 reference reactant state structures the RMSDs are 0.78 kcal/mol
and 0.93 kcal/mol, respectively, for the two reactions. As the inset
demonstrates, the minimum energy path (MEP) is well reproduced by
these parametrizations, too. Such a quality is comparable with that
found in previous studies. The reactive PES for the hydrogen
abstraction reaction between MgO$^{+}$ and ethane had a RMSD of 1.5
kcal/mol with respect to MP2/aug-cc-pVTZ for the reactant and 1.1
kcal/mol for the product state.\cite{sweeny2020thermal} For the
Diels-Alder reaction between 2,3-dibromobutadiene and maleic anhydride
had a RMSD of 2.9 kcal/mol for the overall
reaction.\cite{rivero2021reactive} For the product side (ion-pair) the
RMSDs along the C--Cl and C--Br separations are 3.60 kcal/mol and 2.41
kcal/mol for the two reactions, respectively, and dominated by high
energy structures for long C--Cl$^{-}$ separations.\\

\subsection{Umbrella Sampling Simulations and Potentials of Mean Force}
\begin{figure}[H]
\begin{center}
\includegraphics[width=\linewidth]{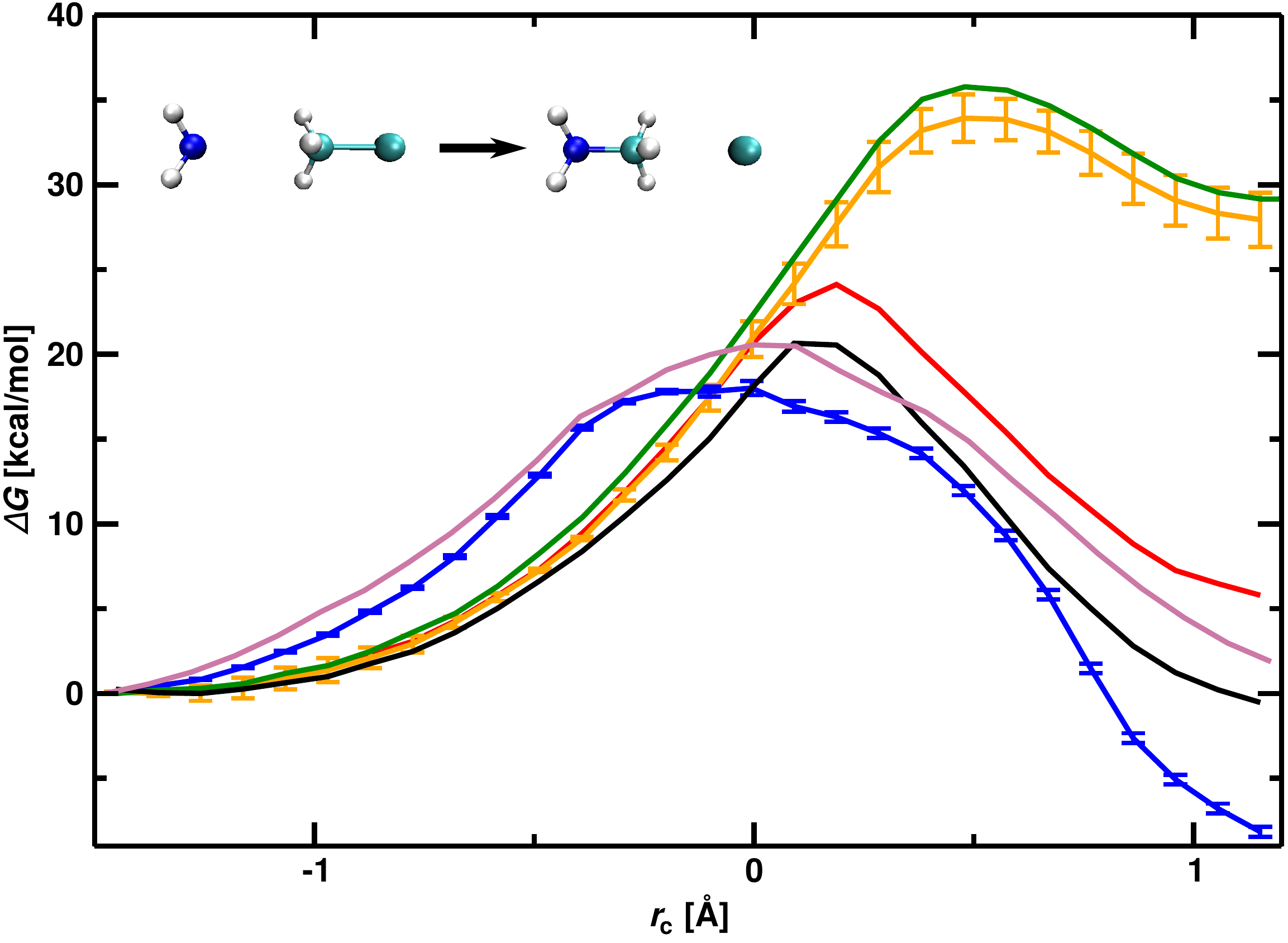}
\caption{Potentials of mean force for the Menshutkin reaction for
  NH$_{3}$+MeCl in water (blue), methanol (purple), benzene (red),
  acetonitrile (black), and cyclohexane (orange) from umbrella
  sampling simulations. For comparison, the PMF from US in the gas
  phase is also reported in green.}
\label{fig:pmf_mecl}
\end{center}
\end{figure}

\textit{NH$_{3}$+MeCl:} The 1-dimensional PMFs for the NH$_{3}$+MeCl
reaction are reported in Figure~\ref{fig:pmf_mecl} and
Table~\ref{tab:tab3} summarizes the activation free barriers.
Umbrella sampling simulations based on the parametrized, reactive
force field for the NH$_{3}$+MeCl reaction yields a free energy
barrier $\Delta G^{\ddagger} = 35.8$ kcal/mol in the gas-phase. This
compares with an activation barrier of 38.5 kcal/mol from the IRC in
the gas-phase at the MP2/6-311++G(2d,2p) level. Previous electronic
structure calculations for the same reaction reported barrier heights
of 36.2 kcal/mol and 32.7 kcal/mol at the
HF/6-31++G*\cite{poater2001effects} and B3PW91/6-31+G*
\cite{halls2002chemistry} levels, respectively. \\

\begin{table}[H]
\centering
\begin{tabular}{l|c|c|c|c|c|c}
		& Gas	& Water & Methanol & Acetonitrile & Benzene & Hexane \\
\hline
Sim. &35.8&18.0 $\pm$ 0.5&20.5&20.6&24.1&33.9 $\pm$ 1.4 \\
Exp.	&       &   23.5\cite{okamoto1967kinetic2}    &   20.8\cite{okamoto1967kinetic}       &       &         & \\
Literature & \begin{tabular}{@{}c@{}} 36.2\cite{poater2001effects} /\\ 32.7 \cite{halls2002chemistry} \end{tabular}	& \begin{tabular}{@{}c@{}}  16.1\cite{poater2001effects}/ 21.7\cite{Castejon1999} / 25.6\cite{gao1991priori} / \\ $15.7 \pm 0.3$\cite{pan2019accelerated} / $21.9 \pm 2.7$\cite{pan2019accelerated} \end{tabular} &         &\begin{tabular}{@{}c@{}}16.8\cite{tang2020solvent} / \\ 22.3\cite{tang2020solvent}\end{tabular}&   &    32.6\cite{Castejon1999}  \\
\end{tabular}
\caption{Free energy barrier for the Menshutkin reaction of
  NH$_{3}$+MeCl (in kcal/mol) calculated with MS-ARMD and US in
  different solvents (Sim.). The experimental (Exp.)  $\Delta
  G^{\ddagger}$ value is for the NH$_3$ + MeI reaction in water (0.05
  \textsc{m} MeI)\cite{okamoto1967kinetic2} and in methanol (0.1
  \textsc{m} MeI).\cite{okamoto1967kinetic} Hexane = cyclohexane.}
\label{tab:tab3}
\end{table}

\noindent
In water, methanol, acetonitrile, benzene, and cyclohexane the barrier
heights from the 1-dimensional PMFs are $18.0 \pm 0.5$, 20.5, 20.6,
24.1, and $33.9 \pm 1.4$ kcal/mol, respectively. Representative error
bars were determined from bootstrapping for polar (water) and apolar
(cyclohexane) solvent. For the simulations in water and methanol, the
activation energies of $18.0 \pm 0.5$ and 20.6 kcal/mol compare with
the experimentally reported values of 23.5\cite{okamoto1967kinetic2}
and 20.8 kcal/mol\cite{okamoto1967kinetic} for NH$_3$ + MeI,
respectively. Only qualitative comparison between experiment and
simulations is possible because a) results from experiments are only
available for MeI and not MeBr, and b) the MeI concentration was 0.05
\textsc{m} in water and 0.1 \textsc{m} in
methanol.\cite{okamoto1967kinetic2} Previously, the minimum energy
pathway for the NH$_{3}$+MeCl reaction in explicit water was
determined at the HF/6-31G* level of theory, followed by single-point
energy calculations at the MP4SDTQ/6-31+G* level. The resulting
activation barrier was 25.6 kcal/mol.\cite{gao1991priori} In a more
recent study the barrier heights from a semi-empirical QM study were
$15.7 \pm 0.3$ and $21.9 \pm 2.7$ kcal/mol,
respectively.\cite{pan2019accelerated} The activation barrier was
$15.7 \pm 0.3$ kcal/mol when directly computed from the lower level
PM3/MM calculations and increased to $21.9 \pm 2.7$ kcal/mol by using
a higher level, indirect B3LYP/MM methodology. In this method,
configurational sampling and iterative pathway optimization were
carried out with a lower-level PM3/MM Hamiltonian, and analyzed with
WHAM to obtain the lower-level reaction free energy profile. This was
followed by thermodynamic perturbation\cite{zwanzig1954high} to a
higher level target B3LYP/MM Hamiltonian with 6-31G* basis set to
obtain a corrected free energy profile.\cite{pan2019accelerated}\\

\noindent
As expected, polar solvents show a stronger effect in reducing the
reaction barrier height compared to apolar solvents, see Figure
\ref{fig:pmf_mecl}. All polar solvents and benzene find an earlier TS
compared to vacuum for which it is at $r_c = 0.5$ \AA\/. For benzene,
acetonitrile, methanol and water the TS progressively shifts towards
the reactant with $r_c$ ranging from $r_c = 0.2$ \AA\/ to $r_c = -0.1$
\AA\/. Furthermore, the free energy profile widens considerably when
going from the gas phase to methanol and water as a solvent and
compared with acetonitrile and the apolar solvents. Water and methanol
are the smallest and most polar solvents and are able to approach and
interact with the solute more directly than the larger and apolar
solvent molecules which leads to widening of the PMF. For benzene the
packing is easier due to its planar structure compared to the chair
conformation of cyclohexane. Further, benzene has stronger
electrostatic interactions with the solute as a result of negatively
charged carbon atoms. Thus, benzene displays a barrier reduction
closer to the polar water solvent than the apolar cyclohexane.\\

\begin{figure}[H]
\begin{center}
\includegraphics[width=0.7\linewidth]{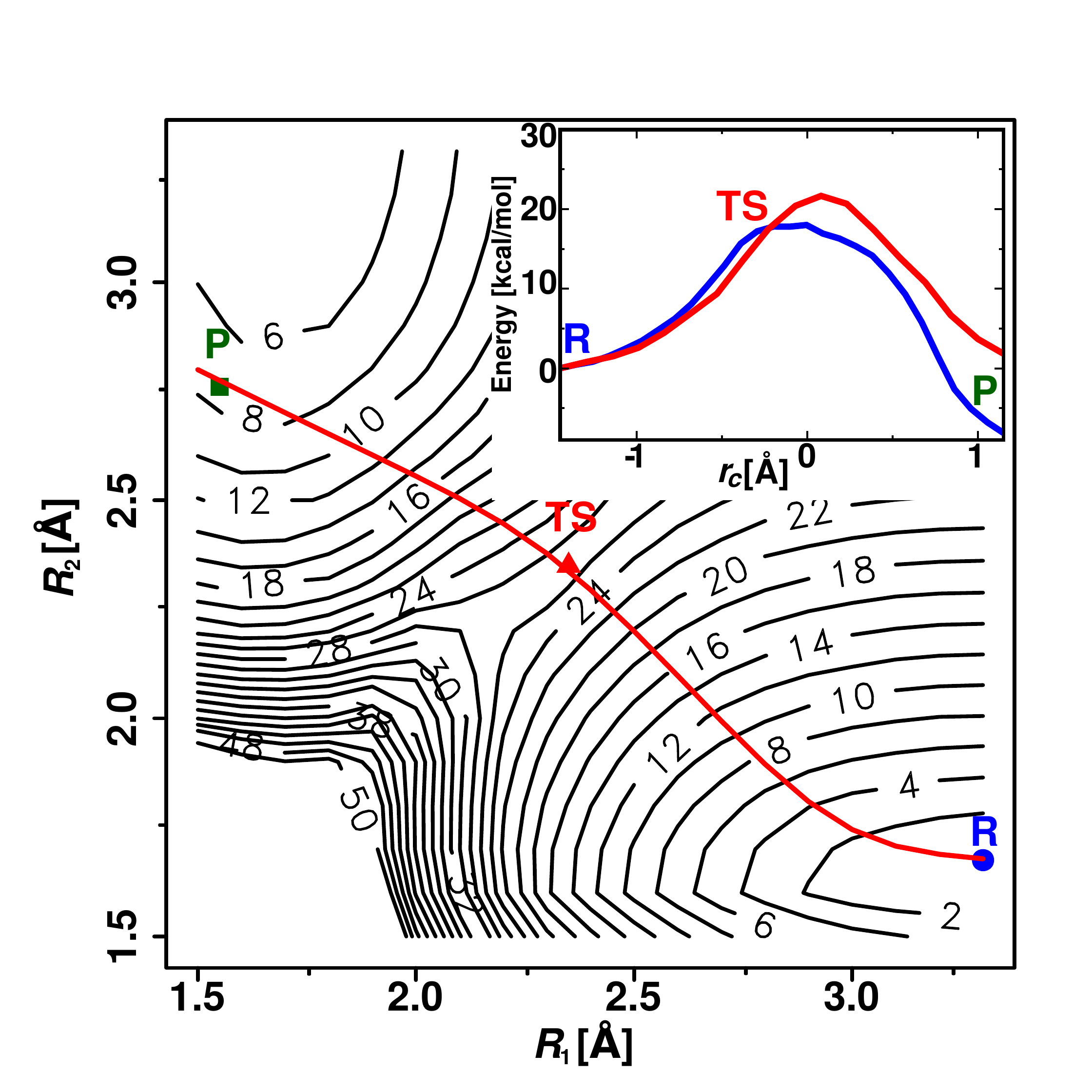}
\caption{2D PMF of NH$_3$+MeCl in water, scanned through reaction
  coordinates of $R_1$ = $d_{\rm C-N}$ and $R_2$ = $d_{\rm
    C-Cl}$. Relative positions of reactant (blue circle), transition
  state (red triangle) and product (green square) shown on the
  plot. The red line denotes the minimum energy pathway from reactant
  to product (see text). Contours are drawn in increments of 2
  kcal/mol.  The inset shows the 1D PMF extracted from the 2D surface
  (red line) superimposed onto the 1D PMF (blue line) of NH$_{3}$+MeCl
  in water from Figure~\ref{fig:pmf_mecl}.}
\label{fig:2d_water}
\end{center}
\end{figure}

\noindent
In order to assess how representative the 1-dimensional PMFs are, a
2-dimensional PMF surface was determined for NH$_{3}$+MeCl in water,
see Figure~\ref{fig:2d_water}. The two reaction coordinates $R_1$ and
$R_2$ were the distance between the methyl-C and ammonia-N atom ($R_1
= d_{\rm C-N}$) and between the methyl-C and the Cl atoms ($R_2 =
d_{\rm C-Cl}$), respectively. In the simulations, $R_1$ and $R_2$ were
constrained by a harmonic potential with a force constant of 1000
kcal/mol and the N--C--Cl angle was constrained to remain linear. For
a fixed $R_1$ value the 2D PMF was scanned along the $R_2$ coordinate
and statistics accumulated from 50 ps sampling for each value of the
reaction coordinate. From this, the 2d PMF was constructed using
2D-WHAM\cite{Kumar1992,souaille.comphycom.2001.us} with a tolerance of
0.001. \\

\noindent
A direct comparison between the 1D PMF from the US (blue) and the
minimum energy path on the 2D PMF (red) is shown in the inset of
Figure \ref{fig:2d_water}. The minimum energy path on the 2D PMF was
calculated by fixing $R_2$ values on a grid with $\Delta R_2 = 0.1$
\AA\/ and minimizing along $R_1$. This provides a realistic path
between reactant and TS. Between TS and the product the true MEP is
slightly different which is, however, of lesser interest in the
present work. The activation barrier calculated from the 2D PMF is
21.7 kcal/mol which is 3.7 kcal/mol higher than that of the 1D PMF
(see Figure~\ref{fig:2d_water} inset) and closer to the experimentally
determined activation barrier of 23.5 kcal/mol for the NH$_3$ + MeI
reaction. The fact that the 1D and 2D PMF are quite close to one
another supports the results of the 1D PMFs reported for the other
solvents.\cite{okamoto1967kinetic2} Part of the difference in the
barrier heights between the 1D and 2D PMFs can be attributed to the
constraints that were used in computing the 2D PMF. On the other hand,
the 2D PMF yields additional insights into mechanistic aspects of the
reaction. The PES clarifies that the Menshutkin reaction is concerted
from the perspective of the two reaction coordinates. As $R_1$
contracts from the reactant (R) to the product (P), $R_2$
elongates. The transition region is comparatively steep along the
transition pathway but flat orthogonal to it. For separations $R_1$
and $R_2$ smaller than 2.2 \AA\/ a strong repulsion is found. \\

\begin{figure}[H]
\begin{center}
\includegraphics[width=\linewidth]{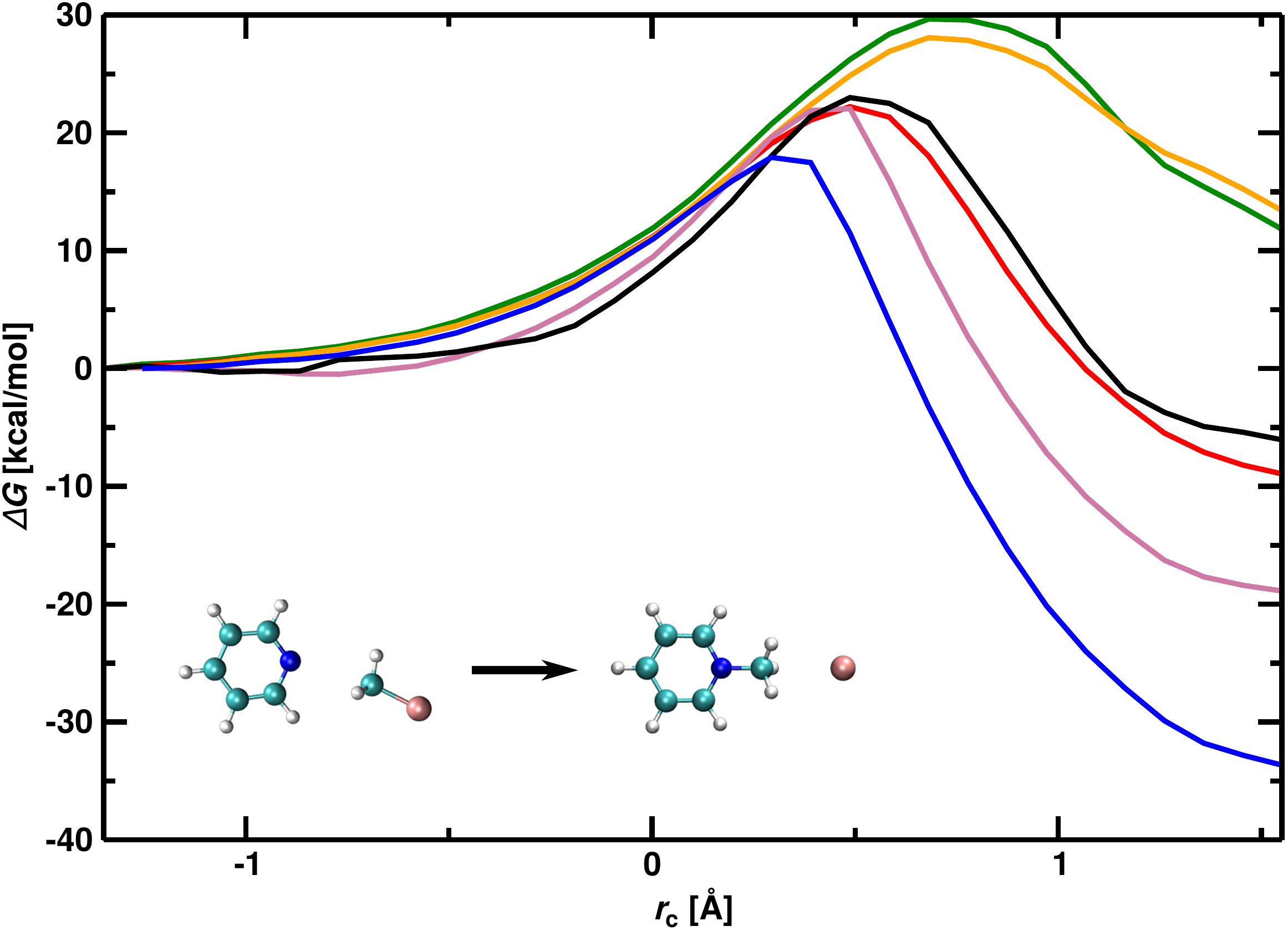}
\caption{Potentials of mean force for the Menshutkin reaction for
  Pyr+MeBr in water (blue), methanol (purple), acetonitrile (black),
  benzene (red), and cyclohexane (orange). For comparison, the result
  from US in the gas phase is reported in green. }
\label{fig:pmf_mebr}
\end{center}
\end{figure}

\noindent
\textit{Pyr+MeBr:} The 1-dimensional PMFs for the Pyr+MeBr Menshutkin
reaction are reported in Figure \ref{fig:pmf_mebr} and Table
\ref{tab:tab2} summarizes the activation free barriers. In the gas
phase US yields an activation free energy of $\Delta G^{\ddagger} =
29.7$ kcal/mol compared with the barrier height of 29.0 kcal/mol from
the IRC at the MP2/6-311++G(2d,2p) level of theory. In water,
methanol, and acetonitrile the computed barriers are 17.9, 22.1, and
23.2 kcal/mol and in apolar solvents they are 22.2 and 28.1 kcal/mol
for benzene and hexane, respectively. Experimentally reported barrier
heights as determined from the measured rates for acetonitrile and
hexane were 22.5 and 27.6 kcal/mol, respectively,\cite{Castejon1999}
and the ordering and absolute values agree favourably with the present
simulations.\\

\begin{table}[H]
\centering
\begin{tabular}{l|c|c|c|c|c|c}
		& Gas	& Water & Methanol & Acetonitrile & Benzene & Hexane \\
\hline
Sim. & 29.7&17.9 &22.1 & 23.2 &22.2 &28.1 \\
Exp.\cite{Castejon1999}	&       &       &          & 22.5        &   &  27.6 \\
\end{tabular}
\caption{Free energy barrier for the Menshutkin reaction of Pyr+MeBr
  (in kcal/mol) calculated with MS-ARMD and US in different solvents
  (Sim.). The experimental (Exp.)  values for $\Delta G^{\ddagger}$
  are from Ref.\cite{Castejon1999} for Pyr+MeBr. Hexane =
  cyclohexane.}
\label{tab:tab2}
\end{table}

\noindent
For polar solvents the reaction barrier height decreases with
increasing polarity of the solvent. Notably, for the reaction in
benzene, the barrier height of 22.2 kcal/mol is close to that for
methanol and lower than that for acetonitrile. The catalytic effect of
benzene has been observed previously from computations for the
Menshutkin reaction of NH$_{3}$ + NH$_{2}$−C$_{6}$H$_{4}$-Br
$\rightarrow$ NH$_{2}$-C$_{6}$H$_{4}$-NH$_{3}^{+}$ + Br$^{-}$ at the
MP2/6-31+G(d) level\cite{Jiang2013}. In this case, the gas-phase
activation barrier of 33.2 kcal/mol decreased to 22.8 kcal/mol in
benzene.\\

\noindent
The position of the TS in the different solvents is directly
correlated with the catalytic effect. Cyclohexane, which displays a
lower barrier to that in the gas phase has the TS at $r_c \sim 0.7$
\AA\/, whereas solvents with catalytic effect shift the TS to
progressively shorter values of $r_c$. Benzene, which reduces the
barrier height slightly more strongly than methanol, has a TS at
around $r_c= 0.5$ \AA\/ which is also the value for methanol. In
water, which shows the strongest reduction in the barrier, the maximum
of the PMF is shifted towards an even smaller value of $r_c = 0.3$
\AA\/. This difference in catalytic strength and shift in TS position
can be explained by the stabilizing effects of the three polar
solvents water, methanol and acetonitrile. The aromatic benzene, even
though apolar, achieves a similar stabilization due to its planar
structure and electrostatics.\\

\subsection{Solvent Distributions}
The MD simulations also offer the opportunity to analyze the solvent
distribution along the reaction at molecular resolution. For this,
separate simulations were carried out for the reactant, product, and
transition states of both reactions in all solvents. The solute was
constrained at the optimized geometries of the reactant, TS, and
ion-pair structures from the MP2/6-311G++(2d,2p) calculations and the
solvent was sampled from unbiased simulations.\\

\begin{figure}[H]
\begin{center}
\includegraphics[width=\linewidth]{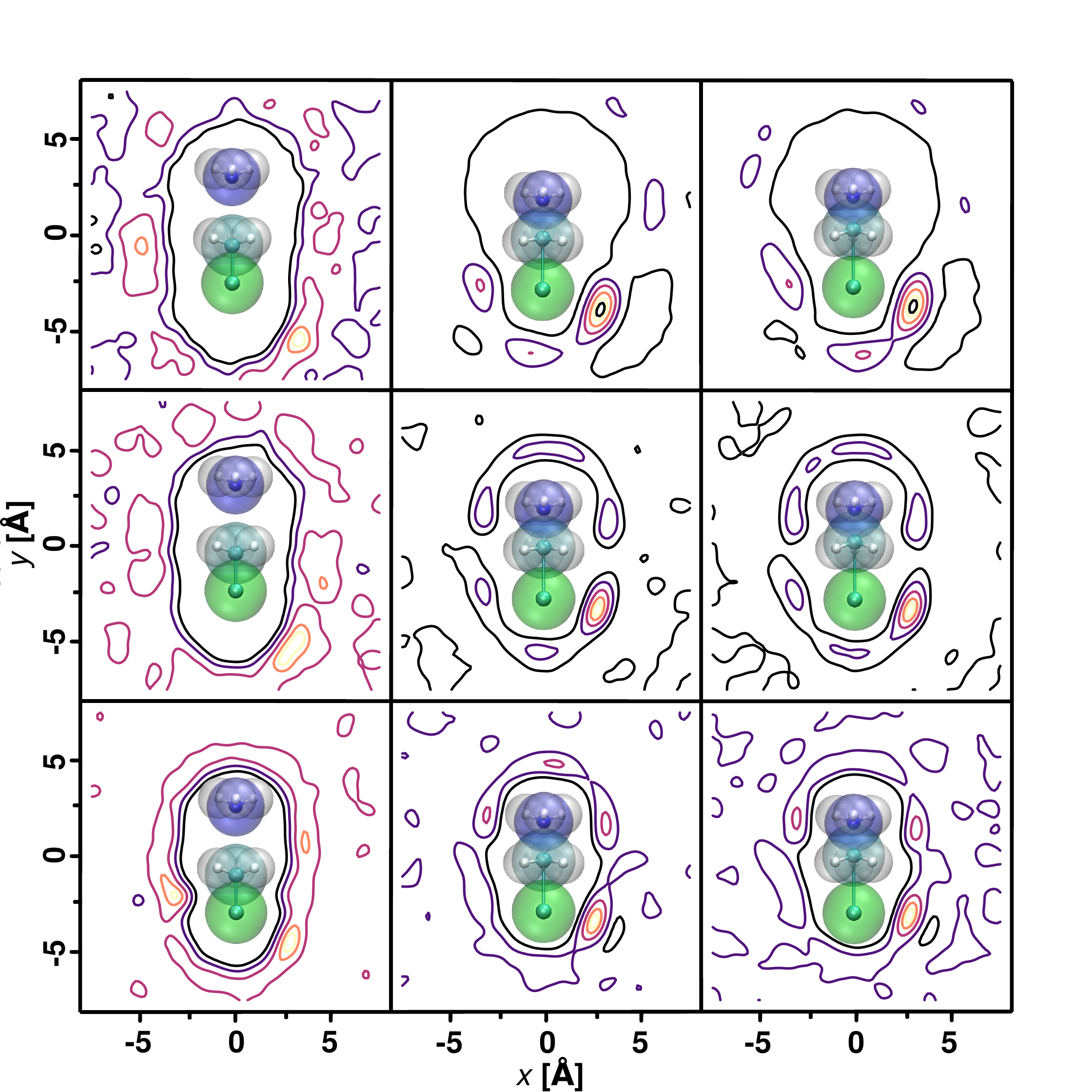}
\caption{2-dimensional solvent distributions for NH$_3$+MeCl from 2 ns
  simulations in polar solvents: acetonitrile, methanol and water
  (from top to bottom) around the reactant, TS, and product state
  structures of the solute (from left to right) projected onto the
  $xy-$plane containing the chloride, carbon and nitrogen atoms. Units
  in \AA\/. In the simulations and the figure the solute is in its
  optimized structure for the reactant, TS, and product state,
  respectively, at the MP2/6-311++G(2d,2p) level of theory. The color
  code for atoms is H (white), C (cyan), N (blue) and Cl (green). Note
  the larger solvent rearrangement between reactant and TS compared
  with TS and product structures.}
\label{fig:mecl_polar}
\end{center}
\end{figure}

\begin{figure}[H]
\begin{center}
\includegraphics[width=\linewidth]{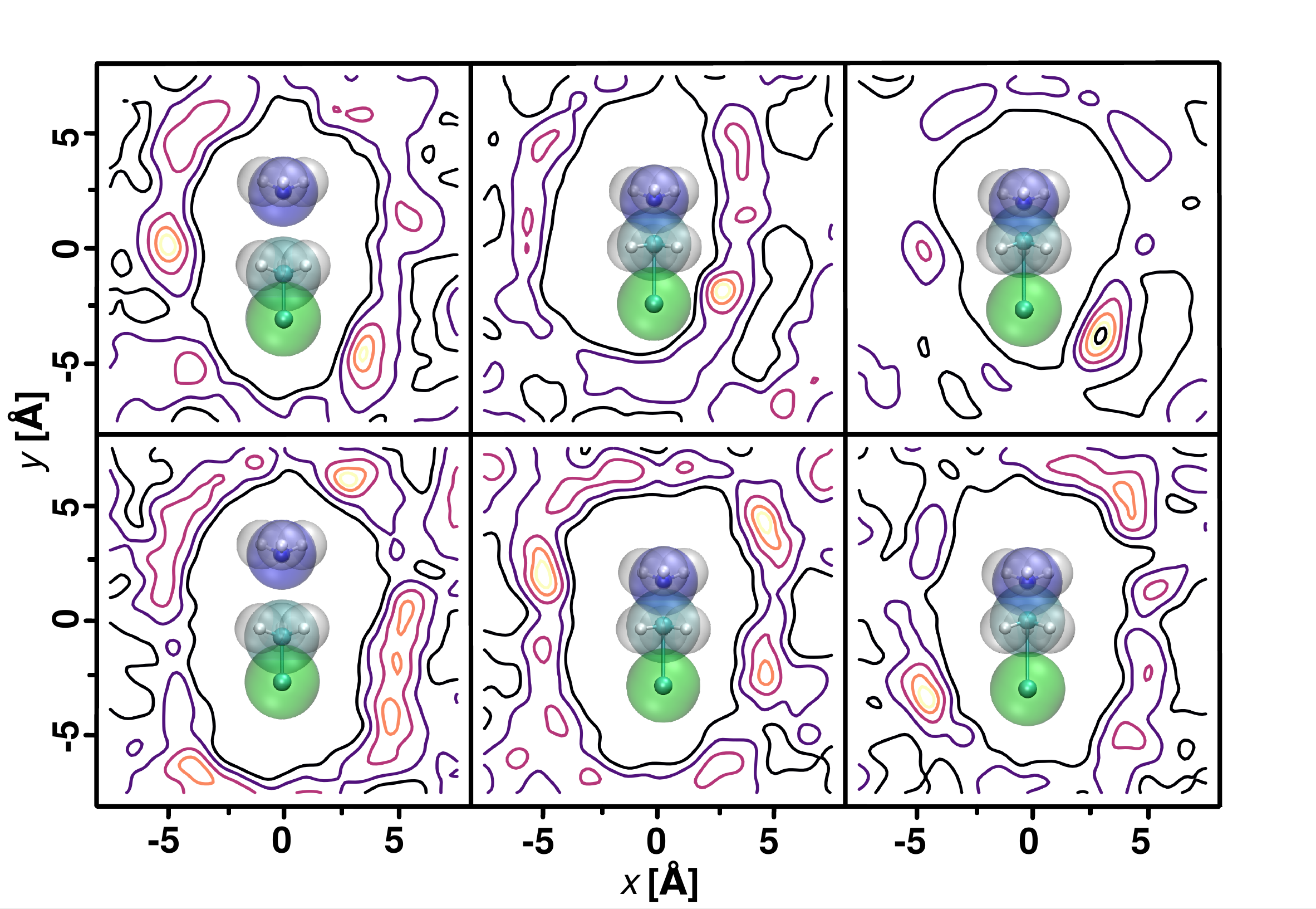}
\caption{2-dimensional solvent distributions for NH$_3$+MeCl in apolar
  solvents: benzene (top) and hexane (bottom) around the reactant, TS,
  and product state structures of the solute (from left to right)
  projected onto the $xy-$plane containing the chloride, carbon an
  nitrogen atoms. Units in \AA\/. In the simulations and the figure
  the solute is in its optimized structure for the reactant, TS, and
  product state, respectively, at the MP2/6-311++G(2d,2p) level of
  theory. The color code for atoms is H (white), C (cyan), N (blue)
  and Cl (green).}
\label{fig:mecl_apolar}
\end{center}
\end{figure}

\noindent
For the NH$_3$+MeCl reaction the solvent distribution around the
solute in polar (water, methanol, acetonitrile) and apolar (benzene
and hexane) solvent is presented in Figures~\ref{fig:mecl_polar}
and~\ref{fig:mecl_apolar}, respectively. In general, for polar
solvents the distribution around the reactant and the TS differs. This
is most evident for acetonitrile and methanol
(Figure~\ref{fig:mecl_polar} top and middle) but also for water
(Figures~\ref{fig:mecl_polar} bottom). Solvent density maxima around
the Cl$^{-}$ anion in the TS were found for methanol, water, and
acetonitrile. The asymmetry around Cl$^{-}$ anion is partly due to the
thin analysis slab (--1\AA\/ $< z <$ 1\AA\/) which was necessary to
avoid congestion of the solvent distribution for other parts of the
molecule. The solvent density maxima are explained by the favourable
interaction of the OH-hydrogen atoms (water and methanol) or the
positively charged carbon atom (acetonitrile) with the Cl$^{-}$
anion. For the TS in methanol and water there are three density maxima
around the H-atoms of ammonia, which are potential H-bonding
sites. For the reactants, the density maximum is around Cl$^{-}$ in
polar solvents. For all three structures, the innermost isocontour is
closest to the solute for water as solvent as packing is tightest due
to its small size.\\

\noindent
Solvent structures for the TS and the product state are
comparable. This can, in part, be rationalized by the similarity of
the TS and ion-pair structures of the solute and the differences
between the reactant and TS structures: the N-C distances for the
Cl--Me--NH$_3$ arrangement are 3.30 \AA\/, 1.81 \AA\/, and 1.55 \AA\/
for the reactant, TS, and ion-pair in vacuum, respectively. Similarly,
the C-Cl distances are 1.80 \AA\/, 2.42 \AA\/, and 2.76 \AA\/ for
reactant, TS and ion-pair. This similarity between the TS and ion-pair
structure is reminiscent of a manifestation of the Hammond postulate
which states that if two states occur consecutively along a reaction
and have nearly the same energy content, their interconversion will
involve only a small reorganization of the molecular structures. For
the solute structures this is evidently the case as the TS and the
contact-ion-pair structures only differ little. Similarly, the solvent
distributions between reactant and TS differ, in general, considerably
more compared with the change in solvent structure between TS and the
product.\\

\noindent
For the apolar solvents, the distribution around the solute showed
less pronounced maxima, see Figure~\ref{fig:mecl_apolar}, except for
the product state in benzene. For cyclohexane the solvent distribution
is considerably less structured than for the polar solvents which
indicates the weak interaction between solute and solvent in this case
and is the underlying reason for the limited catalytic effect of
cyclohexane for the NH$_{3}$+MeCl reaction. Contrary to cyclohexane,
the density maximum of solvent molecules around the Cl$^{-}$ anion for
the product in benzene, and to some degree for the TS, were
present. The packing around the solute was more dense in benzene,
especially at the TS. \\

\noindent
For the Pyr+MeBr reaction the solvent distribution around the solute
in polar and apolar solvents is presented in
Figures S1 and S2,
respectively. Similar to the NH$_{3}$+MeCl reaction, the solvent
distribution around the reactant and the TS differed. For the reactant
the N$_{\rm Pyr}$-C-Br angle deviates from linearity whereas in the TS
and for the contact-ion-pair it was linear. Due to the positively
charged pyridine-H atoms multiple density maxima around the ring were
observed for all states in polar solvents. In methanol, the solvent
distribution was heterogeneous, with multiple density maxima around
the solute and beyond the first solvation shell. Compared with
acetonitrile, methanol is more tightly packed around the
solute. Contrary to that, the solvent distribution for water was
comparatively homogeneous. Although multiple density maxima for water
near the solute are observed for the first solvation shell, the
solvent distribution becomes more homogeneous beyond the first
solvation shell for all three states of the solute. For the
contact-ion-pair solvent structuring occurs for water and methanol,
but not for acetonitrile.\\

\noindent
For apolar solvents, see Figure S2, the solvent
distribution around the solute was less dense compared to the polar
solvents. The maxima around the positively charged hydrogen atoms
appear for benzene. However, the densities around the Br$^{-}$ anion
were less pronounced than for polar solvents. The packing around the
Br$^{-}$ anion was tighter for the TS and contact-ion-pair compared
with the reactant state. Cyclohexane packs more densely around the
solute for Pyr+MeBr compared to NH$_{3}$+MeCl. This can be explained
by the larger size of the solute, which allows the bulkier, apolar
solvent to accommodate better, than around a smaller solute. The trend
of similar solvent distribution around TS and product is also found
for these two solvents.  \\

\subsection{Solvent Energetics}
The simulations also provide quantitative information about the
relative solvent-solvent interactions along the reaction pathway with
respect to an arbitrary reference state which is the reactant in the
following. Two different analyses were carried out for the energetics
of solvent molecules as a function of the reaction coordinate. First,
the energetics of the solvent molecules within the first solvation
shell, i.e. molecules in direct contact with the solute, were
investigated as a function of the reaction coordinate from the
umbrella sampling simulations. Secondly, a similar analysis was
carried out from 2 ns simulations with the solute frozen in the
reactant, TS, and contact-ion-pair geometries at the
MP2/6-311G++(2d,2p) level of theory for a) the first solvation shell
and b) for the entire simulation box.\\

\begin{figure}[H]
\begin{center}
\includegraphics[width=0.8\linewidth]{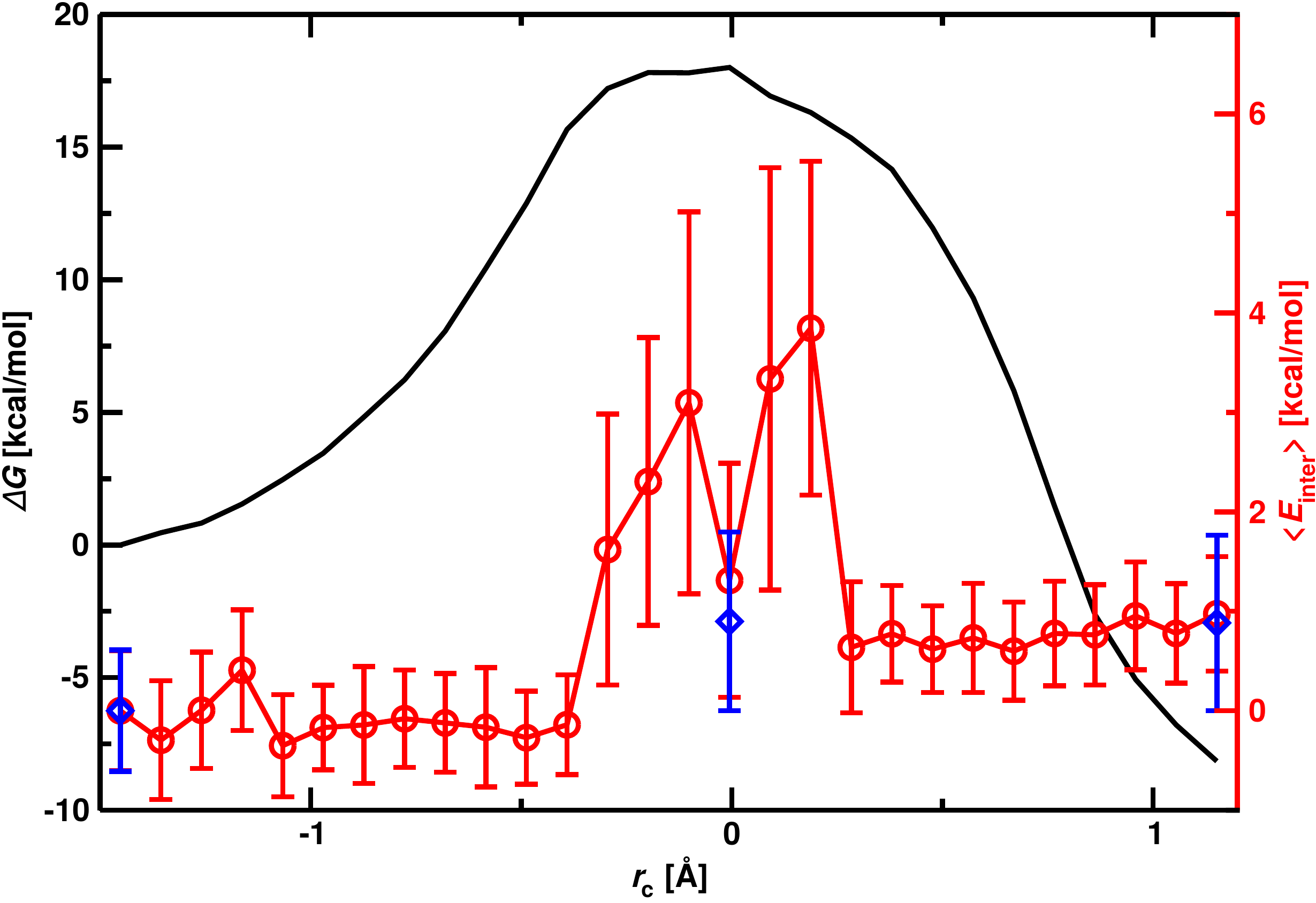}
\caption{Potentials of mean force for the Menshutkin reaction for
  NH$_{3}$+MeCl in water (black line) and the energy of solvent
  molecules within 5 \AA\/ of the solute in kcal/mol/molecule (red
  line). The circles are the mean for a given umbrella together with
  the standard deviation from the mean as a bar. The blue diamonds are
  the average solvent interaction energies from the 2 ns simulations
  with the frozen solute (reactant, TS and contact-ion-pair)
  structures together with the standard deviation from the mean as a
  bar.}
\label{fig:small_box_MeCl}
\end{center}
\end{figure}

\noindent
For the first analysis only solvent molecules within a cutoff distance
of 5 \AA\/ of any atom of the solute were retained which corresponds
approximately to the first solvation shell for each solvent. For each
of the umbrellas from the US simulations the average solvent-solvent
interaction energy per solvent molecule was determined together with
the fluctuation around the mean for the NH$_3$+MeCl reaction in water
(see Figure \ref{fig:small_box_MeCl}) and for the Pyr+MeBr reaction in
methanol (see Figure S3).\\

\noindent
The average energy per solvent molecule with respect to the reactant
within 5 \AA\/ (red trace) superimposed on the PMF of NH$_{3}$+MeCl in
water (black) is shown in Figure~\ref{fig:small_box_MeCl}. Starting
from the reactant structure at $r_c = -1.2$ \AA\/ the average
interaction energy remains around $\sim -0.5$ kcal/mol/molecule up to
$r_c \sim -0.4$ \AA\/ after which it suddenly increases. Before the
transition state the average interaction energy per water molecule is
destabilized by $\sim 2.5$ kcal/mol to accommodate the transition
state of the solute. At the transition state, $r_c = 0$, the solvent
is destabilized with respect to the reactant by about 1 kcal/mol per
water molecule (but stabilized relative to the structures before the
TS by about 1.5 kcal/mol). Beyond the TS the surrounding water
molecules arrange again in a more unfavourable conformation before
their average interaction energies return to levels comparable to that
of the reactant of around $\sim 0.8$ kcal/mol at $r_{c} \sim 0.3$
\AA\/. This demonstrates that the water solvent participates actively
in the reaction progress.\\

\noindent
For the Pyr+MeBr reaction in methanol the average energy per solvent
molecule with respect to the reactant within 5 \AA\/ is shown in
Figure S3. Starting from the reactant
structure the average energy per solvent molecule remains at $\sim -1$
kcal/mol/molecule up to $r_{c} \sim -0.8$ \AA\/. Upon approaching the
transition state ($r_{c} \sim -0.5$ \AA\/) the average interaction
energy between the solvent molecules increases
(destabilization). Compared with the NH$_{3}$+MeCl reaction in water
the destabilization of the solvent starts earlier, though. Before
reaching the TS at $r_{c} \sim 0.4$ \AA\/ the average interaction
energy decreases and reaches a value similar to that of the reactant
albeit with larger variance. Between the TS and the product structure
the interaction energy increases again and then drops back to levels
of the reactant structure. Since the TS and product states are
structurally similar to each other (see Figure S1
and its discussion), the effects on the average solvent energetics are
less prominent in the later stage of the reaction.\\

\noindent
The 2D-solvent distributions for $r_{c} = [-0.3,-0.2,-0.1]$ \AA\/ are
shown in Figure S4. It is found that the
density maxima around MeBr decrease in amplitude as the reaction
proceeds from reactant towards TS, especially for the region between
the methyl-H and the Pyr-H. The protic nature of methanol promotes
H-bonding which results in density maxima around the region between
the methyl-H and the Pyr-H. As the isocontours are drawn at the same
heights in all representations, the populations are directly
comparable. It is found that as the system approaches the region with
$r_{c} = -0.2$ \AA\/ from either the reactant or the TS side, the
ordering decreases appreciably thus, correlated motions of solvent
molecules are required in this region of the reaction profile. This
partly explains in a ``time lapse'' picture why reactions ``take more
time'' (ps and longer) than the actual reactive step (which is rather
fs). which is reflected in the energetic destabilization of the
solvent at this position relative to the reactant or TS
structure. Hence, changes in the solvent distribution are directly
reflected in the average solvent-solvent energetics.\\

\noindent
Again, the more extended 2 ns $NVT$ simulations with the solute frozen
in its reactant, TS, and product state geometry were analyzed and the
average interaction energies for solvent molecules within 5 \AA\/ of
the solute were determined, see Tables~\ref{tab:tab_mecl} and
\ref{tab:tab_mebr}, and the blue diamonds in
Figure S3. These results compare favourably
with the analysis of the US simulations and indicate that the sampling
from the US simulations is representative.\\

\begin{table}[H]
\centering
\caption{Solvent-solvent interaction for NH$_{3}$+MeCl for solvent
  molecules within 5 \AA\/ of any of the solute atoms. The average
  energy per solvent molecule of TS and product states with respect to
  the reactant, in kcal/mol. The standard deviation from the mean per
  solvent molecule given in parentheses. Water-45 labels the
  simulations in the 45 \AA\/ water box.}
\begin{tabular}{l r r r}
\toprule
Solvent          &  Reactant  &  \multicolumn{1}{c}{TS}  &   \multicolumn{1}{c}{Product}  \\
\midrule 
Water            &0 (\footnotesize{0.06}) &  0.06 (\footnotesize{0.06)} & 0.08 (\footnotesize{0.06)} \\
Water-45       & 0 (\footnotesize{0.93)} & 0.07 (\footnotesize{1.05)} & 0.09 (\footnotesize{1.04)} \\
Methanol      &0 (\footnotesize{0.65)} & 0.30 (\footnotesize{0.41)} & 0.41 (\footnotesize{0.32)} \\
Acetonitrile  &0 (\footnotesize{0.28)} & 0.02 (\footnotesize{0.28)} & 0.10 (\footnotesize{0.27)} \\
Benzene       &0 (\footnotesize{2.03)} & 0.67 (\footnotesize{6.30)}& 0.49 (\footnotesize{2.34)}  \\
Cyclohexane & 0 (\footnotesize{1.87)} & 0.21 (\footnotesize{2.26)} & 0.24 (\footnotesize{2.22)} \\
\bottomrule
\label{tab:tab_mecl}
\end{tabular}
\end{table}

\noindent
In addition, one simulation for the NH$_{3}$+MeCl reaction in a 45
\AA\/ cubic box (Water-45 in Table \ref{tab:tab_mecl}) was carried out
to assess the influence of the box size on the solvent
energetics. With an increased number of water molecules the
fluctuation in the solvent interaction energies increases for the
first solvation shell. When analyzing the average interaction energies
and fluctuations around the mean for all water molecules for the two
box sizes the differences were found to be insignificant. Hence, the
box size for the simulations appears to influence the structural
dynamics of the water molecules closest to the solute. This is an
effect worth to be explored but outside the scope of the present
work.\cite{MM.hb:2018} For the solvent-solvent energies a more general
observation is also that with the reactant state as the zero of
energy, the average energy content of the solvent for the TS and
contact-ion pair structures is generally higher.\\

\begin{table}[H]
\centering
\caption{Solvent-solvent interaction for Pyr+MeBr for solvent
  molecules within 5 \AA\/ of any of the solute atoms. The average
  total energy of solvent boxes of TS and product per solvent molecule
  with respect to the reactant, in kcal/mol. The standard deviation
  from the mean per solvent molecule given in parentheses.}
\begin{tabular}{l r r r}
\toprule
Solvent  &  Reactant  &  \multicolumn{1}{c}{TS}  &   \multicolumn{1}{c}{Product}  \\
\midrule
Water &0 \footnotesize{(0.23)} &  0.61 \footnotesize{(0.53)} & 0.62 \footnotesize{(0.55)} \\
Methanol &0 \footnotesize{(0.13)} & 0.40 \footnotesize{(0.39)} & 0.38 \footnotesize{(0.40)} \\
Acetonitrile & 0 \footnotesize{(1.10)} &2.13 \footnotesize{(1.16)}& 2.18 \footnotesize{(1.20)} \\
Benzene &0 \footnotesize{(3.88)} & 0.65 \footnotesize{(4.71)} & 0.70 \footnotesize{(4.64)} \\
Cyclohexane &0 \footnotesize{(1.84)}& 0.19 \footnotesize{(2.15)} & 0.25 \footnotesize{(2.23)}  \\
\bottomrule
\label{tab:tab_mebr}
\end{tabular}
\end{table}

\section{Conclusion}
The energetics and solvent distributions for two Menshutkin reactions
are quantitatively characterized and analyzed at molecular
detail. Barrier height reductions in going from the gas phase to more
polar solvents are consistent with what is known from experiments. The
solvent distributions change appreciably between reactant, transition
state, and product states. These changes in solvent structure are also
reflected in the average solvent-solvent interactions. One notable
feature of the solvent-solvent interactions is the fact that the
fluctuation around the mean increases considerably when going from
small to larger solvent molecules. \\

\noindent
Starting from the reactant structure the average solvent-solvent
interaction within the first solvation shell ($\sim 5$ \AA\/ within
the solute) remains small and constant up to about 50 \% of the
barrier height after which it increases rapidly (see Figures
\ref{fig:small_box_MeCl} and S3). Around the
transition state the solvent-solvent strain relaxes, increases again
and then returns to levels slightly higher than that of the reactant
state. This points towards an intimate interplay between solute and
solvent degrees of freedom along the reaction coordinate. It is also
of interest to note that the fluctuations around the mean increase
appreciably around the TS. Analysis of the solvent distributions also
suggests that in approaching the TS collective motions of the solvent
molecules are required. This may be one of the reasons why time scales
for reactions can differ dramatically from the actual time to cross
the barrier. It will be of interest to compare these findings with
those from unbiased simulations. However, to obtain sufficient
statistics a large number of reactive trajectories will be required,
and ideally a system with a lower activation barrier is considered.\\

\noindent
In summary, a computationally tractable and qualitatively correct
description - as compared with the few experimental data available -
of the Menshutkin reaction the gas phase and in various solvents has
been presented. Analysis of the solvent degrees of freedom point
towards tight coupling between solute and solvent dynamics with
increased fluctuations in the solvent-solvent interactions around the
transition state. The necessary solvent reorganization between
reactant and TS structures of the solvent requires extensive sampling
which is reflected in the time scale separation between the true rate
of a reaction in solution and the time required to traverse the
barrier.\\

\section*{Data Availability}
Data sets are available from the corresponding author upon reasonable
request.

\section*{Supporting Information}
The Supporting Information contains Tables S1 to
S4 for the parametrization of the force fields and
additional Figures S1 to
S4.

\section{Acknowledgments}
The authors gratefully acknowledge financial support from the Swiss
National Science Foundation through grant 200021-117810 and to the
NCCR-MUST.  \\

\bibliography{references}

\providecommand{\latin}[1]{#1}
\makeatletter
\providecommand{\doi}
  {\begingroup\let\do\@makeother\dospecials
  \catcode`\{=1 \catcode`\}=2 \doi@aux}
\providecommand{\doi@aux}[1]{\endgroup\texttt{#1}}
\makeatother
\providecommand*\mcitethebibliography{\thebibliography}
\csname @ifundefined\endcsname{endmcitethebibliography}
  {\let\endmcitethebibliography\endthebibliography}{}
\begin{mcitethebibliography}{88}
\providecommand*\natexlab[1]{#1}
\providecommand*\mciteSetBstSublistMode[1]{}
\providecommand*\mciteSetBstMaxWidthForm[2]{}
\providecommand*\mciteBstWouldAddEndPuncttrue
  {\def\EndOfBibitem{\unskip.}}
\providecommand*\mciteBstWouldAddEndPunctfalse
  {\let\EndOfBibitem\relax}
\providecommand*\mciteSetBstMidEndSepPunct[3]{}
\providecommand*\mciteSetBstSublistLabelBeginEnd[3]{}
\providecommand*\EndOfBibitem{}
\mciteSetBstSublistMode{f}
\mciteSetBstMaxWidthForm{subitem}{(\alph{mcitesubitemcount})}
\mciteSetBstSublistLabelBeginEnd
  {\mcitemaxwidthsubitemform\space}
  {\relax}
  {\relax}

\bibitem[Warshel and Weiss(1980)Warshel, and Weiss]{Warshel1980}
Warshel,~A.; Weiss,~R.~M. {An Empirical Valence Bond Approach for Comparing
  Reactions in Solutions and in Enzymes}. \emph{J. Am. Chem. Soc.}
  \textbf{1980}, \emph{102}, 6218--6226\relax
\mciteBstWouldAddEndPuncttrue
\mciteSetBstMidEndSepPunct{\mcitedefaultmidpunct}
{\mcitedefaultendpunct}{\mcitedefaultseppunct}\relax
\EndOfBibitem
\bibitem[Yadav \latin{et~al.}(1991)Yadav, Jackson, Holbrook, and
  Warshel]{Yadav1991}
Yadav,~A.; Jackson,~R.~M.; Holbrook,~J.~J.; Warshel,~A. {Role of Solvent
  Reorganization Energies in the Catalytic Activity of Enzymes}. \emph{J. Am.
  Chem. Soc.} \textbf{1991}, \emph{113}, 4800--4805\relax
\mciteBstWouldAddEndPuncttrue
\mciteSetBstMidEndSepPunct{\mcitedefaultmidpunct}
{\mcitedefaultendpunct}{\mcitedefaultseppunct}\relax
\EndOfBibitem
\bibitem[Luzhkov and Warshel(1991)Luzhkov, and Warshel]{Luzhkov1991}
Luzhkov,~V.; Warshel,~A. {Microscopic Calculations of Solvent Effects on
  Absorption Spectra of Conjugated Molecules}. \emph{J. Am. Chem. Soc.}
  \textbf{1991}, \emph{113}, 4491--4499\relax
\mciteBstWouldAddEndPuncttrue
\mciteSetBstMidEndSepPunct{\mcitedefaultmidpunct}
{\mcitedefaultendpunct}{\mcitedefaultseppunct}\relax
\EndOfBibitem
\bibitem[Zhan \latin{et~al.}(2017)Zhan, M{\aa}rtensson, Purg, Kamerlin, and
  Ahlquist]{Zhan2017}
Zhan,~S.; M{\aa}rtensson,~D.; Purg,~M.; Kamerlin,~S.~C.; Ahlquist,~M.~S.
  {Capturing the Role of Explicit Solvent in the Dimerization of Ru$^V$(bda)
  Water Oxidation Catalysts}. \emph{Angew. Chemie - Int. Ed.} \textbf{2017},
  \emph{56}, 6962--6965\relax
\mciteBstWouldAddEndPuncttrue
\mciteSetBstMidEndSepPunct{\mcitedefaultmidpunct}
{\mcitedefaultendpunct}{\mcitedefaultseppunct}\relax
\EndOfBibitem
\bibitem[Kamerlin \latin{et~al.}(2009)Kamerlin, Haranczyk, and
  Warshel]{Kamerlin2009}
Kamerlin,~S.~C.; Haranczyk,~M.; Warshel,~A. {Are Mixed Explicit/Implicit
  Solvation Models Reliable for Studying Phosphate Hydrolysis? A Comparative
  Study of Continuum, Explicit and Mixed Solvation Models}. \emph{ChemPhysChem}
  \textbf{2009}, \emph{10}, 1125--1134\relax
\mciteBstWouldAddEndPuncttrue
\mciteSetBstMidEndSepPunct{\mcitedefaultmidpunct}
{\mcitedefaultendpunct}{\mcitedefaultseppunct}\relax
\EndOfBibitem
\bibitem[Ranaghan \latin{et~al.}(2004)Ranaghan, Ridder, Szefczyk, Sokalski,
  Hermann, and Mulholland]{Ranaghan2004}
Ranaghan,~K.~E.; Ridder,~L.; Szefczyk,~B.; Sokalski,~W.~A.; Hermann,~J.~C.;
  Mulholland,~A.~J. {Transition State Stabilization and Substrate Strain in
  Enzyme Catalysis: ab initio QM/MM Modelling of the Chorismate Mutase
  Reaction.} \emph{Org. Biomol. Chem.} \textbf{2004}, \emph{2}, 968--980\relax
\mciteBstWouldAddEndPuncttrue
\mciteSetBstMidEndSepPunct{\mcitedefaultmidpunct}
{\mcitedefaultendpunct}{\mcitedefaultseppunct}\relax
\EndOfBibitem
\bibitem[Shaw \latin{et~al.}(2010)Shaw, Woods, and Mulholland]{Shaw2010}
Shaw,~K.~E.; Woods,~C.~J.; Mulholland,~A.~J. {Compatibility of Quantum Chemical
  Methods and Empirical (MM) Water Models in Quantum Mechanics/Molecular
  Mechanics Liquid Water Simulations}. \emph{J. Phys. Chem. Lett.}
  \textbf{2010}, \emph{1}, 219--223\relax
\mciteBstWouldAddEndPuncttrue
\mciteSetBstMidEndSepPunct{\mcitedefaultmidpunct}
{\mcitedefaultendpunct}{\mcitedefaultseppunct}\relax
\EndOfBibitem
\bibitem[Carlsson and {\AA}qvist(2006)Carlsson, and {\AA}qvist]{Carlsson2006}
Carlsson,~J.; {\AA}qvist,~J. {Calculations of Solute and Solvent Entropies from
  Molecular Dynamics Simulations}. \emph{Phys. Chem. Chem. Phys.}
  \textbf{2006}, \emph{8}, 5385--5395\relax
\mciteBstWouldAddEndPuncttrue
\mciteSetBstMidEndSepPunct{\mcitedefaultmidpunct}
{\mcitedefaultendpunct}{\mcitedefaultseppunct}\relax
\EndOfBibitem
\bibitem[Bingemann \latin{et~al.}(2000)Bingemann, King, and
  Crim]{Bingemann2000}
Bingemann,~D.; King,~A.~M.; Crim,~F.~F. {Transient Electronic Absorption of
  Vibrationally Excited CH$_2$I$_2$: Watching Energy Flow in Solution}.
  \emph{J. Chem. Phys.} \textbf{2000}, \emph{113}, 5018--5025\relax
\mciteBstWouldAddEndPuncttrue
\mciteSetBstMidEndSepPunct{\mcitedefaultmidpunct}
{\mcitedefaultendpunct}{\mcitedefaultseppunct}\relax
\EndOfBibitem
\bibitem[Alml{\"{o}}f \latin{et~al.}(2007)Alml{\"{o}}f, Carlsson, and
  {\AA}qvist]{Almlof2007}
Alml{\"{o}}f,~M.; Carlsson,~J.; {\AA}qvist,~J. {Improving the Accuracy of the
  Linear Interaction Energy Method for Solvation Free Energies}. \emph{J. Chem.
  Theory Comput.} \textbf{2007}, \emph{3}, 2162--2175\relax
\mciteBstWouldAddEndPuncttrue
\mciteSetBstMidEndSepPunct{\mcitedefaultmidpunct}
{\mcitedefaultendpunct}{\mcitedefaultseppunct}\relax
\EndOfBibitem
\bibitem[Elles \latin{et~al.}(2004)Elles, Cox, and Crim]{Elles2004}
Elles,~C.~G.; Cox,~M.~J.; Crim,~F.~F. {Vibrational Relaxation of CH$_3$I in the
  Gas Phase and in Solution}. \emph{J. Chem. Phys.} \textbf{2004}, \emph{120},
  6973--6979\relax
\mciteBstWouldAddEndPuncttrue
\mciteSetBstMidEndSepPunct{\mcitedefaultmidpunct}
{\mcitedefaultendpunct}{\mcitedefaultseppunct}\relax
\EndOfBibitem
\bibitem[Preston \latin{et~al.}(2013)Preston, Shaloski, and Crim]{Preston2013}
Preston,~T.~J.; Shaloski,~M.~A.; Crim,~F.~F. {Probing the Photoisomerization of
  CHBr$_3$ and CHI$_3$ in Solution with Transient Vibrational and Electronic
  Spectroscopy}. \emph{J. Phys. Chem. A} \textbf{2013}, \emph{117},
  2899--2907\relax
\mciteBstWouldAddEndPuncttrue
\mciteSetBstMidEndSepPunct{\mcitedefaultmidpunct}
{\mcitedefaultendpunct}{\mcitedefaultseppunct}\relax
\EndOfBibitem
\bibitem[Rivera \latin{et~al.}(2011)Rivera, Winter, Harper, Benjamin, and
  Bradforth]{Rivera2011}
Rivera,~C.~A.; Winter,~N.; Harper,~R.~V.; Benjamin,~I.; Bradforth,~S.~E. {The
  Dynamical Role of Solvent on the ICN Photodissociation Reaction: Connecting
  Experimental Observables Directly with Molecular Dynamics Simulations.}
  \emph{Phys. Chem. Chem. Phys.} \textbf{2011}, \emph{13}, 8269--8283\relax
\mciteBstWouldAddEndPuncttrue
\mciteSetBstMidEndSepPunct{\mcitedefaultmidpunct}
{\mcitedefaultendpunct}{\mcitedefaultseppunct}\relax
\EndOfBibitem
\bibitem[Orr-Ewing(2014)]{Orr-Ewing2014}
Orr-Ewing,~A.~J. {Perspective: Bimolecular Chemical Reaction Dynamics in
  Liquids}. \emph{J. Chem. Phys.} \textbf{2014}, \emph{140}, 090901\relax
\mciteBstWouldAddEndPuncttrue
\mciteSetBstMidEndSepPunct{\mcitedefaultmidpunct}
{\mcitedefaultendpunct}{\mcitedefaultseppunct}\relax
\EndOfBibitem
\bibitem[Claeyssens \latin{et~al.}(2011)Claeyssens, Ranaghan, Lawan, Macrae,
  Manby, Harvey, and Mulholland]{Claeyssens2011}
Claeyssens,~F.; Ranaghan,~K.~E.; Lawan,~N.; Macrae,~S.~J.; Manby,~F.~R.;
  Harvey,~J.~N.; Mulholland,~A.~J. {Analysis of Chorismate Mutase Catalysis by
  QM/MM Modelling of Enzyme-Catalysed and Uncatalysed Reactions.} \emph{Org.
  Biomol. Chem.} \textbf{2011}, \emph{9}, 1578--1590\relax
\mciteBstWouldAddEndPuncttrue
\mciteSetBstMidEndSepPunct{\mcitedefaultmidpunct}
{\mcitedefaultendpunct}{\mcitedefaultseppunct}\relax
\EndOfBibitem
\bibitem[Severance and Jorgensen(1992)Severance, and
  Jorgensen]{severance.jacs.1992.ave}
Severance,~D.~L.; Jorgensen,~W.~L. {Effects of Hydration on the Claisen
  Rearrangement of Allyl Vinyl Ether from Computer Simulations}. \emph{J. Am.
  Chem. Soc.} \textbf{1992}, \emph{114}, 10966--10968\relax
\mciteBstWouldAddEndPuncttrue
\mciteSetBstMidEndSepPunct{\mcitedefaultmidpunct}
{\mcitedefaultendpunct}{\mcitedefaultseppunct}\relax
\EndOfBibitem
\bibitem[Guest \latin{et~al.}(1997)Guest, Craw, Vincent, and
  Hillier]{guest.perk2.1997.ave}
Guest,~J.~M.; Craw,~J.~S.; Vincent,~M.~A.; Hillier,~I.~H. {The Effect of Water
  on the Claisen Rearrangement of Allyl Vinyl Ether: Theoretical Methods
  Including Explicit Solvent and Electron Correlation}. \emph{Perkin Trans. 2}
  \textbf{1997}, 71--74\relax
\mciteBstWouldAddEndPuncttrue
\mciteSetBstMidEndSepPunct{\mcitedefaultmidpunct}
{\mcitedefaultendpunct}{\mcitedefaultseppunct}\relax
\EndOfBibitem
\bibitem[Cramer and Truhlar(1992)Cramer, and Truhlar]{Cramer1992}
Cramer,~C.~J.; Truhlar,~D.~G. {What Causes Aqueous Acceleration of the Claisen
  Rearrangement?} \emph{J. Am. Chem. Soc.} \textbf{1992}, \emph{114},
  8794--8799\relax
\mciteBstWouldAddEndPuncttrue
\mciteSetBstMidEndSepPunct{\mcitedefaultmidpunct}
{\mcitedefaultendpunct}{\mcitedefaultseppunct}\relax
\EndOfBibitem
\bibitem[Brickel and Meuwly(2019)Brickel, and Meuwly]{MM.claisen:2019}
Brickel,~S.; Meuwly,~M. Molecular Determinants for Rate Acceleration in the
  Claisen Rearrangement Reaction. \emph{J. Phys. Chem. B} \textbf{2019},
  \emph{123}, 448--456\relax
\mciteBstWouldAddEndPuncttrue
\mciteSetBstMidEndSepPunct{\mcitedefaultmidpunct}
{\mcitedefaultendpunct}{\mcitedefaultseppunct}\relax
\EndOfBibitem
\bibitem[Hwang \latin{et~al.}(1988)Hwang, King, Creighton, and
  Warshel]{Hwang1988}
Hwang,~J.~K.; King,~G.; Creighton,~S.; Warshel,~A. {Simulation of Free Energy
  Rrelationships and Dynamics of S$_N$2 Reactions in Aqueous Solution}.
  \emph{J. Am. Chem. Soc.} \textbf{1988}, \emph{110}, 5297--5311\relax
\mciteBstWouldAddEndPuncttrue
\mciteSetBstMidEndSepPunct{\mcitedefaultmidpunct}
{\mcitedefaultendpunct}{\mcitedefaultseppunct}\relax
\EndOfBibitem
\bibitem[Shaik(1985)]{Shaik1985}
Shaik,~S. {The Collage of S$_N$2 Reactivity Patterns: A State Correlation
  Diagram Model}. \emph{Prog. Phys. Org. Chem.} \textbf{1985}, \emph{15},
  197--337\relax
\mciteBstWouldAddEndPuncttrue
\mciteSetBstMidEndSepPunct{\mcitedefaultmidpunct}
{\mcitedefaultendpunct}{\mcitedefaultseppunct}\relax
\EndOfBibitem
\bibitem[Chandrasekhar \latin{et~al.}(1985)Chandrasekhar, Smith, and
  Jorgensen]{Chandrasekhar1985}
Chandrasekhar,~J.; Smith,~S.~F.; Jorgensen,~W.~L. {Theoretical Examination of
  the S$_N$2 Reaction Involving Chloride Ion and Methyl Chloride in the Gas
  Phase and Aqueous Solution}. \emph{J. Am. Chem. Soc.} \textbf{1985},
  \emph{107}, 154--163\relax
\mciteBstWouldAddEndPuncttrue
\mciteSetBstMidEndSepPunct{\mcitedefaultmidpunct}
{\mcitedefaultendpunct}{\mcitedefaultseppunct}\relax
\EndOfBibitem
\bibitem[Merkel \latin{et~al.}(1988)Merkel, Zahradn{\'{i}}k, and
  Havlas]{Merkel1988}
Merkel,~A.; Zahradn{\'{i}}k,~R.; Havlas,~Z. {Evaluation of the Rate Constant
  for the S$_N$2 Reaction CH$_3$F + H$^-$ $\rightarrow$ CH$_4$ + F$^-$ in the
  Gas Phase}. \emph{J. Am. Chem. Soc.} \textbf{1988}, \emph{110},
  8355--8359\relax
\mciteBstWouldAddEndPuncttrue
\mciteSetBstMidEndSepPunct{\mcitedefaultmidpunct}
{\mcitedefaultendpunct}{\mcitedefaultseppunct}\relax
\EndOfBibitem
\bibitem[Gao and Xia(1993)Gao, and Xia]{Gao1993}
Gao,~J.; Xia,~X. {A Two-Dimensional Energy Surface for a type II S$_N$2
  Reaction in Aqueous Solution}. \emph{J. Am. Chem. Soc.} \textbf{1993},
  \emph{115}, 9667--9675\relax
\mciteBstWouldAddEndPuncttrue
\mciteSetBstMidEndSepPunct{\mcitedefaultmidpunct}
{\mcitedefaultendpunct}{\mcitedefaultseppunct}\relax
\EndOfBibitem
\bibitem[Fradera \latin{et~al.}(1996)Fradera, Amat, Torrent, Mestres, Constans,
  Besald, Marti, Simon, Lobato, Oliva, Luis, Sol, Carbb, and
  Duran]{Fradera1996}
Fradera,~X.; Amat,~L.; Torrent,~M.; Mestres,~J.; Constans,~P.; Besald,~E.;
  Marti,~J.; Simon,~S.; Lobato,~M.; Oliva,~J.~M. \latin{et~al.}  {Analysis of
  the Changes on the Potential Energy Surface of Menshutkin Reactions Induced
  by External Perturbations}. \emph{J. Mol. Struct.} \textbf{1996}, \emph{371},
  171--183\relax
\mciteBstWouldAddEndPuncttrue
\mciteSetBstMidEndSepPunct{\mcitedefaultmidpunct}
{\mcitedefaultendpunct}{\mcitedefaultseppunct}\relax
\EndOfBibitem
\bibitem[Adamovic and Gordon(2005)Adamovic, and Gordon]{Adamovic2005}
Adamovic,~I.; Gordon,~M.~S. {Solvent Effects on the S$_N$2 Reaction:
  Application of the Density Functional Theory-Based Effective Fragment
  Potential Method}. \emph{J. Phys. Chem. A} \textbf{2005}, \emph{109},
  1629--1636\relax
\mciteBstWouldAddEndPuncttrue
\mciteSetBstMidEndSepPunct{\mcitedefaultmidpunct}
{\mcitedefaultendpunct}{\mcitedefaultseppunct}\relax
\EndOfBibitem
\bibitem[Shaik \latin{et~al.}(2006)Shaik, Ioffe, Reddy, and Pross]{Shaik2006}
Shaik,~S.; Ioffe,~A.; Reddy,~A.~C.; Pross,~A. {Is the Avoided Crossing State a
  Good Approximation for the Transition State of a Chemical Reaction? An
  Analysis of Menschutkin and Ionic S$_N$2 Reactions}. \emph{J. Am. Chem. Soc.}
  \textbf{2006}, \emph{116}, 262--273\relax
\mciteBstWouldAddEndPuncttrue
\mciteSetBstMidEndSepPunct{\mcitedefaultmidpunct}
{\mcitedefaultendpunct}{\mcitedefaultseppunct}\relax
\EndOfBibitem
\bibitem[Menschutkin(1890)]{Menschutkin1890a}
Menschutkin,~N. {Beitr{\"{a}}ge zur Kenntnis der Affinit{\"{a}}tskoeffizienten
  der Alkylhaloide und der Organischen Amine}. \emph{Zeitschrift f{\"{u}}r
  Phys. Chemie} \textbf{1890}, \emph{5}, 589--600\relax
\mciteBstWouldAddEndPuncttrue
\mciteSetBstMidEndSepPunct{\mcitedefaultmidpunct}
{\mcitedefaultendpunct}{\mcitedefaultseppunct}\relax
\EndOfBibitem
\bibitem[Menschutkin(1890)]{Menschutkin1890}
Menschutkin,~N. {{\"{U}}ber die Affinit{\"{a}}tskoeffizienten der Alkylhaloide
  und der Amine}. \emph{Zeitschrift f{\"{u}}r Phys. Chemie} \textbf{1890},
  \emph{6}, 41--57\relax
\mciteBstWouldAddEndPuncttrue
\mciteSetBstMidEndSepPunct{\mcitedefaultmidpunct}
{\mcitedefaultendpunct}{\mcitedefaultseppunct}\relax
\EndOfBibitem
\bibitem[Allfrey \latin{et~al.}(1964)Allfrey, Faulkner, and
  Mirsky]{allfrey:1964}
Allfrey,~V.~G.; Faulkner,~R.; Mirsky,~A. Acetylation and Methylation of
  Histones and Their Possible Role in the Regulation of RNA Synthesis.
  \emph{Proc. Nat. Acad. Sci. USA} \textbf{1964}, \emph{51}, 786\relax
\mciteBstWouldAddEndPuncttrue
\mciteSetBstMidEndSepPunct{\mcitedefaultmidpunct}
{\mcitedefaultendpunct}{\mcitedefaultseppunct}\relax
\EndOfBibitem
\bibitem[Schubert \latin{et~al.}(2003)Schubert, Blumenthal, and
  Cheng]{Schubert2003}
Schubert,~H.~L.; Blumenthal,~R.~M.; Cheng,~X. {Many Paths to Methyltransfer: A
  Chronicle of Convergence}. \emph{Trends Biochem. Sci.} \textbf{2003},
  \emph{28}, 329--335\relax
\mciteBstWouldAddEndPuncttrue
\mciteSetBstMidEndSepPunct{\mcitedefaultmidpunct}
{\mcitedefaultendpunct}{\mcitedefaultseppunct}\relax
\EndOfBibitem
\bibitem[Schmidt \latin{et~al.}(2014)Schmidt, Schwede, and Meuwly]{Schmidt2014}
Schmidt,~T.; Schwede,~T.; Meuwly,~M. {Computational Analysis of Methyl Transfer
  Reactions in Dengue Virus Methyltransferase}. \emph{J. Phys. Chem. B}
  \textbf{2014}, \emph{118}, 5882--5890\relax
\mciteBstWouldAddEndPuncttrue
\mciteSetBstMidEndSepPunct{\mcitedefaultmidpunct}
{\mcitedefaultendpunct}{\mcitedefaultseppunct}\relax
\EndOfBibitem
\bibitem[Arava and Diesendruck(2017)Arava, and
  Diesendruck]{arava2017strategies}
Arava,~S.; Diesendruck,~C.~E. Strategies for the Synthesis of N-Arylammonium
  Salts. \emph{Synthesis} \textbf{2017}, \emph{49}, 3535--3545\relax
\mciteBstWouldAddEndPuncttrue
\mciteSetBstMidEndSepPunct{\mcitedefaultmidpunct}
{\mcitedefaultendpunct}{\mcitedefaultseppunct}\relax
\EndOfBibitem
\bibitem[Ding \latin{et~al.}(2015)Ding, Smulan, Hou, Taubert, Watts, and
  Walker]{ding2015s}
Ding,~W.; Smulan,~L.~J.; Hou,~N.~S.; Taubert,~S.; Watts,~J.~L.; Walker,~A.~K.
  S-Adenosylmethionine Levels Govern Innate Immunity Through Distinct
  Methylation-Dependent Pathways. \emph{Cell Metab.} \textbf{2015}, \emph{22},
  633--645\relax
\mciteBstWouldAddEndPuncttrue
\mciteSetBstMidEndSepPunct{\mcitedefaultmidpunct}
{\mcitedefaultendpunct}{\mcitedefaultseppunct}\relax
\EndOfBibitem
\bibitem[Scavetta \latin{et~al.}(2000)Scavetta, Thomas, Walsh, Szegedi,
  Joachimiak, Gumport, and Churchill]{scavetta2000structure}
Scavetta,~R.~D.; Thomas,~C.~B.; Walsh,~M.~A.; Szegedi,~S.; Joachimiak,~A.;
  Gumport,~R.~I.; Churchill,~M.~E. Structure of Rsr I Methyltransferase, a
  Member of the N 6-adenine $\beta$ Class of DNA Methyltransferases.
  \emph{Nucleic Acids Res.} \textbf{2000}, \emph{28}, 3950--3961\relax
\mciteBstWouldAddEndPuncttrue
\mciteSetBstMidEndSepPunct{\mcitedefaultmidpunct}
{\mcitedefaultendpunct}{\mcitedefaultseppunct}\relax
\EndOfBibitem
\bibitem[Tajima \latin{et~al.}(2016)Tajima, Suetake, Takeshita, Nakagawa, and
  Kimura]{tajima2016domain}
Tajima,~S.; Suetake,~I.; Takeshita,~K.; Nakagawa,~A.; Kimura,~H. Domain
  Structure of the Dnmt1, Dnmt3a, and Dnmt3b DNA Methyltransferases. \emph{DNA
  Methyltransferases-Role and Function} \textbf{2016}, 63--86\relax
\mciteBstWouldAddEndPuncttrue
\mciteSetBstMidEndSepPunct{\mcitedefaultmidpunct}
{\mcitedefaultendpunct}{\mcitedefaultseppunct}\relax
\EndOfBibitem
\bibitem[Li \latin{et~al.}(2021)Li, Chen, and Lu]{li2021interplay}
Li,~Y.; Chen,~X.; Lu,~C. The Interplay Between DNA and Histone Methylation:
  Molecular Mechanisms and Disease Implications. \emph{EMBO Rep.}
  \textbf{2021}, e51803\relax
\mciteBstWouldAddEndPuncttrue
\mciteSetBstMidEndSepPunct{\mcitedefaultmidpunct}
{\mcitedefaultendpunct}{\mcitedefaultseppunct}\relax
\EndOfBibitem
\bibitem[Shi \latin{et~al.}(2004)Shi, Lan, Matson, Mulligan, Whetstine, Cole,
  Casero, and Shi]{shi:2004}
Shi,~Y.; Lan,~F.; Matson,~C.; Mulligan,~P.; Whetstine,~J.~R.; Cole,~P.~A.;
  Casero,~R.~A.; Shi,~Y. Histone Demethylation Mediated by the Nuclear Amine
  Oxidase Homolog LSD1. \emph{Cell} \textbf{2004}, \emph{119}, 941--953\relax
\mciteBstWouldAddEndPuncttrue
\mciteSetBstMidEndSepPunct{\mcitedefaultmidpunct}
{\mcitedefaultendpunct}{\mcitedefaultseppunct}\relax
\EndOfBibitem
\bibitem[Lieder and Brauman(1975)Lieder, and Brauman]{Lieder1975}
Lieder,~C.~A.; Brauman,~J.~I. {A Technique for Detection of Neutral Products in
  Gas-Phase, Ion-Molecule Reactions}. \emph{Int. J. Mass Spectrom. Ion Phys.}
  \textbf{1975}, \emph{16}, 307--319\relax
\mciteBstWouldAddEndPuncttrue
\mciteSetBstMidEndSepPunct{\mcitedefaultmidpunct}
{\mcitedefaultendpunct}{\mcitedefaultseppunct}\relax
\EndOfBibitem
\bibitem[Olmstead and Brauman(1977)Olmstead, and Brauman]{Olmstead1977}
Olmstead,~W.~N.; Brauman,~J.~I. {Gas-Phase Nucleophilic Displacement
  Reactions}. \emph{J. Am. Chem. Soc.} \textbf{1977}, \emph{99},
  4219--4228\relax
\mciteBstWouldAddEndPuncttrue
\mciteSetBstMidEndSepPunct{\mcitedefaultmidpunct}
{\mcitedefaultendpunct}{\mcitedefaultseppunct}\relax
\EndOfBibitem
\bibitem[Hierl \latin{et~al.}(1987)Hierl, Ahrens, Henchman, Viggiano, and
  Paulson]{Hierl1987}
Hierl,~P.~M.; Ahrens,~A.~F.; Henchman,~M.; Viggiano,~A.~A.; Paulson,~J.~F.
  {Rate Constants and Product Distributions as Functions of Temperature for the
  Reaction of OH$^-$(H$_2$O)$_{0,1,2}$ with CH$_3$CN}. \emph{Int. J. Mass
  Spectrom. Ion Process.} \textbf{1987}, \emph{81}, 101--122\relax
\mciteBstWouldAddEndPuncttrue
\mciteSetBstMidEndSepPunct{\mcitedefaultmidpunct}
{\mcitedefaultendpunct}{\mcitedefaultseppunct}\relax
\EndOfBibitem
\bibitem[Barlow \latin{et~al.}(1988)Barlow, {Van Doren}, and
  Bierbaum]{Barlow1988}
Barlow,~S.~E.; {Van Doren},~J.~M.; Bierbaum,~V.~M. {The Gas-Phase Displacement
  Reaction of Chloride Ion with Methyl Chloride as a Function of Kinetic
  Energy}. \emph{J. Am. Chem. Soc.} \textbf{1988}, \emph{110}, 7240--7242\relax
\mciteBstWouldAddEndPuncttrue
\mciteSetBstMidEndSepPunct{\mcitedefaultmidpunct}
{\mcitedefaultendpunct}{\mcitedefaultseppunct}\relax
\EndOfBibitem
\bibitem[{Vande Linde} and Hase(1989){Vande Linde}, and Hase]{VandeLinde1989}
{Vande Linde},~S.~R.; Hase,~W.~L. {A Direct Mechanism for S$_N$2 Nucleophilic
  Substitution Enhanced by Mode Selective Vibrational Excitation}. \emph{J. Am.
  Chem. Soc.} \textbf{1989}, \emph{111}, 2349--2351\relax
\mciteBstWouldAddEndPuncttrue
\mciteSetBstMidEndSepPunct{\mcitedefaultmidpunct}
{\mcitedefaultendpunct}{\mcitedefaultseppunct}\relax
\EndOfBibitem
\bibitem[DePuy \latin{et~al.}(1990)DePuy, Gronert, Mullin, and
  Bierbaum]{DePuy1990}
DePuy,~C.~H.; Gronert,~S.; Mullin,~A.; Bierbaum,~V.~M. {Gas-Phase S$_N$2 and E2
  Reactions of Alkyl Halides}. \emph{J. Am. Chem. Soc.} \textbf{1990},
  \emph{112}, 8650--8655\relax
\mciteBstWouldAddEndPuncttrue
\mciteSetBstMidEndSepPunct{\mcitedefaultmidpunct}
{\mcitedefaultendpunct}{\mcitedefaultseppunct}\relax
\EndOfBibitem
\bibitem[Kato \latin{et~al.}(2001)Kato, Davico, Lee, DePuy, and
  Bierbaum]{Kato2001}
Kato,~S.; Davico,~G.~E.; Lee,~H.~S.; DePuy,~C.~H.; Bierbaum,~V.~M. {Deuterium
  Kinetic Isotope Effects in Gas Phase S$_N$2 Reactions}. \emph{Int. J. Mass
  Spectrom.} \textbf{2001}, \emph{210-211}, 223--229\relax
\mciteBstWouldAddEndPuncttrue
\mciteSetBstMidEndSepPunct{\mcitedefaultmidpunct}
{\mcitedefaultendpunct}{\mcitedefaultseppunct}\relax
\EndOfBibitem
\bibitem[Stei \latin{et~al.}(2016)Stei, Carrascosa, Kainz, Kelkar, Meyer,
  Szab{\'{o}}, Czak{\'{o}}, and Wester]{Stei2016}
Stei,~M.; Carrascosa,~E.; Kainz,~M.~A.; Kelkar,~A.~H.; Meyer,~J.;
  Szab{\'{o}},~I.; Czak{\'{o}},~G.; Wester,~R. {Influence of the Leaving Group
  on the Dynamics of a Gas-Phase S$_N$2 Reaction}. \emph{Nat. Chem.}
  \textbf{2016}, \emph{8}, 151--156\relax
\mciteBstWouldAddEndPuncttrue
\mciteSetBstMidEndSepPunct{\mcitedefaultmidpunct}
{\mcitedefaultendpunct}{\mcitedefaultseppunct}\relax
\EndOfBibitem
\bibitem[Carrascosa \latin{et~al.}(2017)Carrascosa, Meyer, Zhang, Stei,
  Michaelsen, Hase, Yang, and Wester]{Carrascosa2017}
Carrascosa,~E.; Meyer,~J.; Zhang,~J.; Stei,~M.; Michaelsen,~T.; Hase,~W.~L.;
  Yang,~L.; Wester,~R. {Imaging Dynamic Fingerprints of Competing E2 and S$_N$2
  Reactions}. \emph{Nat. Commun.} \textbf{2017}, \emph{8}, 1--7\relax
\mciteBstWouldAddEndPuncttrue
\mciteSetBstMidEndSepPunct{\mcitedefaultmidpunct}
{\mcitedefaultendpunct}{\mcitedefaultseppunct}\relax
\EndOfBibitem
\bibitem[Viers \latin{et~al.}(1984)Viers, Schug, Stovall, and
  Seeman]{Viers1984}
Viers,~J.~W.; Schug,~J.~C.; Stovall,~M.~D.; Seeman,~J.~I. MNDO Study of
  Reaction Pathways for S$_{N}$2 Reactions. Menschutkin Reaction Potential
  Energy Surfaces. \emph{J. Comp. Chem.} \textbf{1984}, \emph{5},
  598--605\relax
\mciteBstWouldAddEndPuncttrue
\mciteSetBstMidEndSepPunct{\mcitedefaultmidpunct}
{\mcitedefaultendpunct}{\mcitedefaultseppunct}\relax
\EndOfBibitem
\bibitem[Amovilli \latin{et~al.}(1998)Amovilli, Mennucci, and
  Floris]{Amovilli1998}
Amovilli,~C.; Mennucci,~B.; Floris,~F.~M. MCSCF Study of the S$_{N}$2
  Menshutkin Reaction in Aqueous Solution within the Polarizable Continuum
  Model. \emph{J. Phys. Chem. B} \textbf{1998}, \emph{102}, 3023--3028\relax
\mciteBstWouldAddEndPuncttrue
\mciteSetBstMidEndSepPunct{\mcitedefaultmidpunct}
{\mcitedefaultendpunct}{\mcitedefaultseppunct}\relax
\EndOfBibitem
\bibitem[Castejon \latin{et~al.}(2001)Castejon, Wiberg, Sklenak, and
  Hinz]{Castejon2001}
Castejon,~H.; Wiberg,~K.~B.; Sklenak,~S.; Hinz,~W. {Solvent Effects on Methyl
  Transfer Reactions. 2. The Reaction of Amines with Trimethylsulfonium Salts}.
  \emph{J. Am. Chem. Soc.} \textbf{2001}, \emph{123}, 6092--6097\relax
\mciteBstWouldAddEndPuncttrue
\mciteSetBstMidEndSepPunct{\mcitedefaultmidpunct}
{\mcitedefaultendpunct}{\mcitedefaultseppunct}\relax
\EndOfBibitem
\bibitem[Sola \latin{et~al.}(1991)Sola, Lledos, Duran, Bertran, and
  Abboud]{Sola1991}
Sola,~M.; Lledos,~A.; Duran,~M.; Bertran,~J.; Abboud,~J. L.~M. {Analysis of
  Solvent Effects on the Menshutkin Reaction}. \emph{J. Am. Chem. Soc.}
  \textbf{1991}, \emph{113}, 2873--2879\relax
\mciteBstWouldAddEndPuncttrue
\mciteSetBstMidEndSepPunct{\mcitedefaultmidpunct}
{\mcitedefaultendpunct}{\mcitedefaultseppunct}\relax
\EndOfBibitem
\bibitem[Jiang \latin{et~al.}(2013)Jiang, Orimoto, and Aoki]{Jiang2013}
Jiang,~L.; Orimoto,~Y.; Aoki,~Y. {Substituent Effects on Menshutkin-Type
  Reactions in the Gas Phase and Solutions: Theoretical Approach from the
  Orbital Interaction View}. \emph{J. Chem. Theory Comput.} \textbf{2013},
  \emph{9}, 4035--4045\relax
\mciteBstWouldAddEndPuncttrue
\mciteSetBstMidEndSepPunct{\mcitedefaultmidpunct}
{\mcitedefaultendpunct}{\mcitedefaultseppunct}\relax
\EndOfBibitem
\bibitem[Acevedo and Jorgensen(2010)Acevedo, and Jorgensen]{Acevedo2010}
Acevedo,~O.; Jorgensen,~W.~L. {Exploring Solvent Effects Upon the Menshutkin
  Reaction Using a Polarizable Force Field}. \emph{J. Phys. Chem. B}
  \textbf{2010}, \emph{114}, 8425--8430\relax
\mciteBstWouldAddEndPuncttrue
\mciteSetBstMidEndSepPunct{\mcitedefaultmidpunct}
{\mcitedefaultendpunct}{\mcitedefaultseppunct}\relax
\EndOfBibitem
\bibitem[Castejon and Wiberg(1999)Castejon, and Wiberg]{Castejon1999}
Castejon,~H.; Wiberg,~K.~B. {Solvent Effects on Methyl Transfer Reactions. 1.
  The Menshutkin Reaction}. \emph{J. Am. Chem. Soc.} \textbf{1999}, \emph{121},
  2139--2146\relax
\mciteBstWouldAddEndPuncttrue
\mciteSetBstMidEndSepPunct{\mcitedefaultmidpunct}
{\mcitedefaultendpunct}{\mcitedefaultseppunct}\relax
\EndOfBibitem
\bibitem[Gronet \latin{et~al.}(1991)Gronet, DePuy, and Bierbaum]{Gronet1991}
Gronet,~S.; DePuy,~C.~H.; Bierbaum,~V.~M. {Deuterium Isotope Effects in
  Gas-Phase Reactions of Alkyl Halides: Distinguishing E2 and S$_N$2 Pathways}.
  \emph{J. Am. Chem. Soc.} \textbf{1991}, \emph{113}, 4009--4010\relax
\mciteBstWouldAddEndPuncttrue
\mciteSetBstMidEndSepPunct{\mcitedefaultmidpunct}
{\mcitedefaultendpunct}{\mcitedefaultseppunct}\relax
\EndOfBibitem
\bibitem[Auriel and de~Hoffmann(1979)Auriel, and
  de~Hoffmann]{auriel1979quantitative}
Auriel,~M.; de~Hoffmann,~E. Quantitative Study of Solvent Effects on the
  Menshutkin Reaction Between 1, 4-diazabicyclo [2.2.2] octane and
  2-chloroethylbenzene, 2-bromoethylbenzene, and 2-iodoethylbenzene. Part 2.
  Mixed Solvents. \emph{J. Chem. Soc., Perkin trans. 2} \textbf{1979},
  325--329\relax
\mciteBstWouldAddEndPuncttrue
\mciteSetBstMidEndSepPunct{\mcitedefaultmidpunct}
{\mcitedefaultendpunct}{\mcitedefaultseppunct}\relax
\EndOfBibitem
\bibitem[Maran \latin{et~al.}(1994)Maran, Pakkanen, and
  Karelson]{maran1994semiempirical}
Maran,~U.; Pakkanen,~T.~A.; Karelson,~M. Semiempirical Study of the Solvent
  Effect on the Menshutkin Reaction. \emph{J. Chem. Soc., Perkin trans. 2}
  \textbf{1994}, 2445--2452\relax
\mciteBstWouldAddEndPuncttrue
\mciteSetBstMidEndSepPunct{\mcitedefaultmidpunct}
{\mcitedefaultendpunct}{\mcitedefaultseppunct}\relax
\EndOfBibitem
\bibitem[Komarova \latin{et~al.}(2021)Komarova, Dick, and
  Luterbacher]{komarova2021diformylxylose}
Komarova,~A.~O.; Dick,~G.~R.; Luterbacher,~J.~S. {Diformylxylose as a New Polar
  Aprotic Solvent Produced from Renewable Biomass}. \emph{Green Chem.}
  \textbf{2021}, \emph{23}, 4790--4799\relax
\mciteBstWouldAddEndPuncttrue
\mciteSetBstMidEndSepPunct{\mcitedefaultmidpunct}
{\mcitedefaultendpunct}{\mcitedefaultseppunct}\relax
\EndOfBibitem
\bibitem[Poater \latin{et~al.}(2001)Poater, Sol{\`a}, Duran, and
  Fradera]{poater2001effects}
Poater,~J.; Sol{\`a},~M.; Duran,~M.; Fradera,~X. Effects of Solvation on the
  Pairing of Electrons in a Series of Simple Molecules and in the Menshutkin
  Reaction. \emph{J. Phys. Chem. A} \textbf{2001}, \emph{105}, 6249--6257\relax
\mciteBstWouldAddEndPuncttrue
\mciteSetBstMidEndSepPunct{\mcitedefaultmidpunct}
{\mcitedefaultendpunct}{\mcitedefaultseppunct}\relax
\EndOfBibitem
\bibitem[Halls and Schlegel(2002)Halls, and Schlegel]{halls2002chemistry}
Halls,~M.~D.; Schlegel,~H.~B. Chemistry Inside Carbon Nanotubes: the Menshutkin
  S$_{N}$2 Reaction. \emph{J. Phys. Chem. B} \textbf{2002}, \emph{106},
  1921--1925\relax
\mciteBstWouldAddEndPuncttrue
\mciteSetBstMidEndSepPunct{\mcitedefaultmidpunct}
{\mcitedefaultendpunct}{\mcitedefaultseppunct}\relax
\EndOfBibitem
\bibitem[Miertu{\v{s}} \latin{et~al.}(1981)Miertu{\v{s}}, Scrocco, and
  Tomasi]{miertuvs1981electrostatic}
Miertu{\v{s}},~S.; Scrocco,~E.; Tomasi,~J. Electrostatic Interaction of a
  Solute with a Continuum. A Direct Utilizaion of $ab$ initio Molecular
  Potentials for the Prevision of Solvent Effects. \emph{Chem. Phys.}
  \textbf{1981}, \emph{55}, 117--129\relax
\mciteBstWouldAddEndPuncttrue
\mciteSetBstMidEndSepPunct{\mcitedefaultmidpunct}
{\mcitedefaultendpunct}{\mcitedefaultseppunct}\relax
\EndOfBibitem
\bibitem[Pan \latin{et~al.}(2019)Pan, Li, Ho, Pu, Mei, and
  Shao]{pan2019accelerated}
Pan,~X.; Li,~P.; Ho,~J.; Pu,~J.; Mei,~Y.; Shao,~Y. Accelerated Computation of
  Free Energy Profile at $ab$ initio Quantum Mechanical/Molecular Mechanical
  Accuracy via a Semi-Empirical Reference Potential. II. Recalibrating
  Semi-Empirical Parameters with Force Matching. \emph{Phys. Chem. Chem. Phys.}
  \textbf{2019}, \emph{21}, 20595--20605\relax
\mciteBstWouldAddEndPuncttrue
\mciteSetBstMidEndSepPunct{\mcitedefaultmidpunct}
{\mcitedefaultendpunct}{\mcitedefaultseppunct}\relax
\EndOfBibitem
\bibitem[Tang \latin{et~al.}(2020)Tang, Zhao, Jiang, Xu, Zhao, and
  Tong]{tang2020solvent}
Tang,~W.; Zhao,~J.; Jiang,~P.; Xu,~X.; Zhao,~S.; Tong,~Z. Solvent Effects on
  the Symmetric and Asymmetric S$_N$2 Reactions in the Acetonitrile Solution: A
  Reaction Density Functional Theory Study. \emph{J. Phys. Chem. B}
  \textbf{2020}, \emph{124}, 3114--3122\relax
\mciteBstWouldAddEndPuncttrue
\mciteSetBstMidEndSepPunct{\mcitedefaultmidpunct}
{\mcitedefaultendpunct}{\mcitedefaultseppunct}\relax
\EndOfBibitem
\bibitem[Nagy \latin{et~al.}(2014)Nagy, Reyes, and
  Meuwly]{nagy.jctc.2014.msarmd}
Nagy,~T.; Reyes,~J.~Y.; Meuwly,~M. Multisurface Adiabatic Reactive Molecular
  Dynamics. \emph{J. Chem. Theory. Comput.} \textbf{2014}, \emph{10},
  1366--1375\relax
\mciteBstWouldAddEndPuncttrue
\mciteSetBstMidEndSepPunct{\mcitedefaultmidpunct}
{\mcitedefaultendpunct}{\mcitedefaultseppunct}\relax
\EndOfBibitem
\bibitem[Brooks \latin{et~al.}(2009)Brooks, Brooks, Mackerell, Nilsson,
  Petrella, Roux, Won, Archontis, Bartels, Boresch, Caflisch, Caves, Cui,
  Dinner, Feig, Fischer, Gao, Hodoscek, Im, Kuczera, Lazaridis, Ma,
  Ovchinnikov, Paci, Pastor, Post, Pu, Schaefer, Tidor, Venable, Woodcock, Wu,
  Yang, York, and Karplus]{charmm.prog}
Brooks,~B.~R.; Brooks,~C.~L.,~III; Mackerell,~A.~D.,~Jr.; Nilsson,~L.;
  Petrella,~R.~J.; Roux,~B.; Won,~Y.; Archontis,~G.; Bartels,~C.; Boresch,~S.
  \latin{et~al.}  {CHARMM: The Biomolecular Simulation Program}. \emph{J. Chem.
  Comp.} \textbf{2009}, \emph{30}, 1545--1614\relax
\mciteBstWouldAddEndPuncttrue
\mciteSetBstMidEndSepPunct{\mcitedefaultmidpunct}
{\mcitedefaultendpunct}{\mcitedefaultseppunct}\relax
\EndOfBibitem
\bibitem[Verlet(1967)]{verlet1967computer}
Verlet,~L. Computer Experiments on Classical Fluids. I. Thermodynamical
  Properties of Lennard-Jones Molecules. \emph{Phys. Rev.} \textbf{1967},
  \emph{159}, 98--103\relax
\mciteBstWouldAddEndPuncttrue
\mciteSetBstMidEndSepPunct{\mcitedefaultmidpunct}
{\mcitedefaultendpunct}{\mcitedefaultseppunct}\relax
\EndOfBibitem
\bibitem[Hoover(1985)]{hoover1985canonical}
Hoover,~W.~G. Canonical Dynamics: Equilibrium Phase-Space Distributions.
  \emph{Phys. Rev. A} \textbf{1985}, \emph{31}, 1695--1697\relax
\mciteBstWouldAddEndPuncttrue
\mciteSetBstMidEndSepPunct{\mcitedefaultmidpunct}
{\mcitedefaultendpunct}{\mcitedefaultseppunct}\relax
\EndOfBibitem
\bibitem[VanGunsteren and Berendsen({1977})VanGunsteren, and
  Berendsen]{W_VanGunsteren_MolPhys_1977}
VanGunsteren,~W.; Berendsen,~H. {Algorithms for Macromolecular Dynamics and
  Constraint Dynamics}. \emph{{Mol. Phys.}} \textbf{{1977}}, \emph{{34}},
  {1311--1327}\relax
\mciteBstWouldAddEndPuncttrue
\mciteSetBstMidEndSepPunct{\mcitedefaultmidpunct}
{\mcitedefaultendpunct}{\mcitedefaultseppunct}\relax
\EndOfBibitem
\bibitem[Essmann \latin{et~al.}(1995)Essmann, Perera, Darden, Lee, Pedersen,
  Essmann, Perera, Berkowitz, Darden, Lee, and Pedersen]{Essmann1995}
Essmann,~U.; Perera,~L.; Darden,~M. L.~B.; Lee,~H.; Pedersen,~L.~G.;
  Essmann,~U.; Perera,~L.; Berkowitz,~M.~L.; Darden,~T.; Lee,~H. \latin{et~al.}
   {A Smooth Particle Mesh Ewald Method.} \emph{J. Chem. Phys.} \textbf{1995},
  \emph{103}, 8577--8593\relax
\mciteBstWouldAddEndPuncttrue
\mciteSetBstMidEndSepPunct{\mcitedefaultmidpunct}
{\mcitedefaultendpunct}{\mcitedefaultseppunct}\relax
\EndOfBibitem
\bibitem[Frisch \latin{et~al.}()Frisch, Trucks, Schlegel, Scuseria, Robb,
  Cheeseman, Scalmani, Barone, Mennucci, Petersson, Nakatsuji, Caricato, Li,
  Hratchian, Izmaylov, Bloino, Zheng, Sonnenberg, Hada, Ehara, Toyota, Fukuda,
  Hasegawa, Ishida, Nakajima, Honda, Kitao, Nakai, Vreven, {Montgomery Jr.},
  Peralta, Ogliaro, Bearpark, Heyd, Brothers, Kudin, Staroverov, Kobayashi,
  Normand, Raghavachari, Rendell, Burant, Iyengar, Tomasi, Cossi, Rega, Millam,
  Klene, Knox, Cross, Bakken, Adamo, Jaramillo, Gomperts, Stratmann, Yazyev,
  Austin, Cammi, Pomelli, Ochterski, Martin, Morokuma, Zakrzewski, Voth,
  Salvador, Dannenberg, Dapprich, Daniels, Farkas, Foresman, Ortiz, Cioslowski,
  and Fox]{g09}
Frisch,~M.~J.; Trucks,~G.~W.; Schlegel,~H.~B.; Scuseria,~G.~E.; Robb,~M.~A.;
  Cheeseman,~J.~R.; Scalmani,~G.; Barone,~V.; Mennucci,~B.; Petersson,~G.~A.
  \latin{et~al.}  {Gaussian09 {Revision {\{}D{\}}.01}}\relax
\mciteBstWouldAddEndPuncttrue
\mciteSetBstMidEndSepPunct{\mcitedefaultmidpunct}
{\mcitedefaultendpunct}{\mcitedefaultseppunct}\relax
\EndOfBibitem
\bibitem[Zoete \latin{et~al.}(2011)Zoete, Cuendet, Grosdidier, and
  Michielin]{zoete.jcc.2011.swissparam}
Zoete,~V.; Cuendet,~M.~A.; Grosdidier,~A.; Michielin,~O. {SwissParam: A Fast
  Force Field Generation Tool for Small Organic Molecules}. \emph{J. Comput.
  Chem.} \textbf{2011}, \emph{32}, 2959--2368\relax
\mciteBstWouldAddEndPuncttrue
\mciteSetBstMidEndSepPunct{\mcitedefaultmidpunct}
{\mcitedefaultendpunct}{\mcitedefaultseppunct}\relax
\EndOfBibitem
\bibitem[Brickel \latin{et~al.}(2019)Brickel, Das, Unke, Turan, and
  Meuwly]{Brickel2019}
Brickel,~S.; Das,~A.; Unke,~O.; Turan,~H.; Meuwly,~M. Reactive Molecular
  Dynamics for the [Cl--CH$_3$-Br]$^-$ Reaction in the Gas Phase and in
  Solution: A Comparative Study Using Empirical and Neural Network Force
  Fields. \emph{Electron. Struct.} \textbf{2019}, \emph{1}, 024002\relax
\mciteBstWouldAddEndPuncttrue
\mciteSetBstMidEndSepPunct{\mcitedefaultmidpunct}
{\mcitedefaultendpunct}{\mcitedefaultseppunct}\relax
\EndOfBibitem
\bibitem[Brickel and Meuwly(2017)Brickel, and Meuwly]{Brickel2017}
Brickel,~S.; Meuwly,~M. {OH-Stretching Overtone Induced Dynamics in HSO$_3$F
  from Reactive Molecular Dynamics Simulations}. \emph{J. Phys. Chem. A}
  \textbf{2017}, \emph{121}, 5079--5087\relax
\mciteBstWouldAddEndPuncttrue
\mciteSetBstMidEndSepPunct{\mcitedefaultmidpunct}
{\mcitedefaultendpunct}{\mcitedefaultseppunct}\relax
\EndOfBibitem
\bibitem[Reyes \latin{et~al.}(2016)Reyes, Brickel, Unke, Nagy, and
  Meuwly]{YosaReyes2016}
Reyes,~J.~Y.; Brickel,~S.; Unke,~O.~T.; Nagy,~T.; Meuwly,~M. HSO$_{3}$Cl: a
  Prototype Molecule for Studying OH-stretching Overtone Induced
  Photodissociation. \emph{Phys. Chem. Chem. Phys.} \textbf{2016}, \emph{18},
  6780--6788\relax
\mciteBstWouldAddEndPuncttrue
\mciteSetBstMidEndSepPunct{\mcitedefaultmidpunct}
{\mcitedefaultendpunct}{\mcitedefaultseppunct}\relax
\EndOfBibitem
\bibitem[Nelder and Mead(1965)Nelder, and Mead]{Nelder1965}
Nelder,~J.; Mead,~R. {A Simplex Method for Function Minimization.} \emph{Chem.
  Phys.} \textbf{1965}, \emph{7}, 308--313\relax
\mciteBstWouldAddEndPuncttrue
\mciteSetBstMidEndSepPunct{\mcitedefaultmidpunct}
{\mcitedefaultendpunct}{\mcitedefaultseppunct}\relax
\EndOfBibitem
\bibitem[Kottalam and Case(1988)Kottalam, and Case]{kottalam.jacs.1988.us}
Kottalam,~J.; Case,~D.~A. {Dynamics of Ligand Escape from the Heme Pocket of
  Myoglobin}. \emph{J. Am. Chem. Soc.} \textbf{1988}, \emph{110},
  7690--7697\relax
\mciteBstWouldAddEndPuncttrue
\mciteSetBstMidEndSepPunct{\mcitedefaultmidpunct}
{\mcitedefaultendpunct}{\mcitedefaultseppunct}\relax
\EndOfBibitem
\bibitem[Kumar \latin{et~al.}(1992)Kumar, Rosenberg, Bouzida, Swendsen, and
  Kollman]{Kumar1992}
Kumar,~S.; Rosenberg,~J.~M.; Bouzida,~D.; Swendsen,~R.~H.; Kollman,~P.~A. {The
  Weighted Histogram Analysis Method for Free-energy Calculations on
  Biomolecules. I. The Method}. \emph{J. Comput. Chem.} \textbf{1992},
  \emph{13}, 1011--1021\relax
\mciteBstWouldAddEndPuncttrue
\mciteSetBstMidEndSepPunct{\mcitedefaultmidpunct}
{\mcitedefaultendpunct}{\mcitedefaultseppunct}\relax
\EndOfBibitem
\bibitem[Souaille and Roux(2001)Souaille, and Roux]{souaille.comphycom.2001.us}
Souaille,~M.; Roux,~B. {Extension to the Weighted Histogram Analysis Method:
  Combining Umbrella Sampling with Free Energy Calculations}. \emph{Comput.
  Phys. Commun.} \textbf{2001}, \emph{135}, 40--57\relax
\mciteBstWouldAddEndPuncttrue
\mciteSetBstMidEndSepPunct{\mcitedefaultmidpunct}
{\mcitedefaultendpunct}{\mcitedefaultseppunct}\relax
\EndOfBibitem
\bibitem[Rosenblatt(1956)]{rosenblatt1956remarks}
Rosenblatt,~M. Remarks on Some Nonparametric Estimates of a Density Function.
  \emph{Ann. Math. Stat.} \textbf{1956}, \emph{27}, 832--837\relax
\mciteBstWouldAddEndPuncttrue
\mciteSetBstMidEndSepPunct{\mcitedefaultmidpunct}
{\mcitedefaultendpunct}{\mcitedefaultseppunct}\relax
\EndOfBibitem
\bibitem[Parzen(1962)]{parzen1962estimation}
Parzen,~E. On Estimation of a Probability Density Function and Mode. \emph{Ann.
  Math. Stat.} \textbf{1962}, \emph{33}, 1065--1076\relax
\mciteBstWouldAddEndPuncttrue
\mciteSetBstMidEndSepPunct{\mcitedefaultmidpunct}
{\mcitedefaultendpunct}{\mcitedefaultseppunct}\relax
\EndOfBibitem
\bibitem[Sweeny \latin{et~al.}(2020)Sweeny, Pan, Kassem, Sawyer, Ard, Shuman,
  Viggiano, Brickel, Unke, Upadhyay, \latin{et~al.} others]{sweeny2020thermal}
Sweeny,~B.~C.; Pan,~H.; Kassem,~A.; Sawyer,~J.~C.; Ard,~S.~G.; Shuman,~N.~S.;
  Viggiano,~A.~A.; Brickel,~S.; Unke,~O.~T.; Upadhyay,~M. \latin{et~al.}
  Thermal Activation of Methane by MgO$^{+}$: Temperature Dependent Kinetics,
  Reactive Molecular Dynamics Simulations and Statistical Modeling. \emph{Phys.
  Chem. Chem. Phys.} \textbf{2020}, \emph{22}, 8913--8923\relax
\mciteBstWouldAddEndPuncttrue
\mciteSetBstMidEndSepPunct{\mcitedefaultmidpunct}
{\mcitedefaultendpunct}{\mcitedefaultseppunct}\relax
\EndOfBibitem
\bibitem[Rivero \latin{et~al.}(2021)Rivero, Turan, Meuwly, and
  Willitsch]{rivero2021reactive}
Rivero,~U.; Turan,~H.~T.; Meuwly,~M.; Willitsch,~S. Reactive Atomistic
  Simulations of Diels-Alder Type Reactions: Conformational and Dynamic Effects
  in the Polar Cycloaddition of 2, 3-dibromobutadiene Radical Ions with Maleic
  Anhydride. \emph{Mol. Phys.} \textbf{2021}, \emph{119}, e1825852\relax
\mciteBstWouldAddEndPuncttrue
\mciteSetBstMidEndSepPunct{\mcitedefaultmidpunct}
{\mcitedefaultendpunct}{\mcitedefaultseppunct}\relax
\EndOfBibitem
\bibitem[Okamoto \latin{et~al.}(1967)Okamoto, Fukui, and
  Shingu]{okamoto1967kinetic2}
Okamoto,~K.; Fukui,~S.; Shingu,~H. Kinetic Studies of Bimolecular Nucleophilic
  Substitution. VI. Rates of the Menschtkin Reaction of Methyl Iodide with
  Methylamines and Ammonia in Aqueous Solutions. \emph{Bull. Chem. Soc. of Jpn}
  \textbf{1967}, \emph{40}, 1920--1925\relax
\mciteBstWouldAddEndPuncttrue
\mciteSetBstMidEndSepPunct{\mcitedefaultmidpunct}
{\mcitedefaultendpunct}{\mcitedefaultseppunct}\relax
\EndOfBibitem
\bibitem[Okamoto \latin{et~al.}(1967)Okamoto, Fukui, Nitta, and
  Shingu]{okamoto1967kinetic}
Okamoto,~K.; Fukui,~S.; Nitta,~I.; Shingu,~H. Kinetic Studies of Bimolecular
  Nucleophilic Substitution. VIII. The Effect of Hydroxylic Solvents on the
  Nucleophilicity of Aliphatic Amines in the Menschutkin Reaction. \emph{Bull.
  Chem. Soc. of Jpn} \textbf{1967}, \emph{40}, 2354--2357\relax
\mciteBstWouldAddEndPuncttrue
\mciteSetBstMidEndSepPunct{\mcitedefaultmidpunct}
{\mcitedefaultendpunct}{\mcitedefaultseppunct}\relax
\EndOfBibitem
\bibitem[Gao(1991)]{gao1991priori}
Gao,~J. A Priori Computation of a Solvent-Enhanced S$_N$2 Reaction Profile in
  Water: the Menshutkin Reaction. \emph{J. Am. Chem. Soc.} \textbf{1991},
  \emph{113}, 7796--7797\relax
\mciteBstWouldAddEndPuncttrue
\mciteSetBstMidEndSepPunct{\mcitedefaultmidpunct}
{\mcitedefaultendpunct}{\mcitedefaultseppunct}\relax
\EndOfBibitem
\bibitem[Zwanzig(1954)]{zwanzig1954high}
Zwanzig,~R.~W. High-Temperature Equation of State by a Perturbation Method. I.
  Nonpolar Gases. \emph{J. Chem. Phys.} \textbf{1954}, \emph{22},
  1420--1426\relax
\mciteBstWouldAddEndPuncttrue
\mciteSetBstMidEndSepPunct{\mcitedefaultmidpunct}
{\mcitedefaultendpunct}{\mcitedefaultseppunct}\relax
\EndOfBibitem
\bibitem[El~Hage \latin{et~al.}(2018)El~Hage, Hedin, Gupta, Meuwly, and
  Karplus]{MM.hb:2018}
El~Hage,~K.; Hedin,~F.; Gupta,~P.~K.; Meuwly,~M.; Karplus,~M. Valid molecular
  dynamics simulations of human hemoglobin require a surprisingly large box
  size. \emph{Elife} \textbf{2018}, \emph{7}, e35560\relax
\mciteBstWouldAddEndPuncttrue
\mciteSetBstMidEndSepPunct{\mcitedefaultmidpunct}
{\mcitedefaultendpunct}{\mcitedefaultseppunct}\relax
\EndOfBibitem
\end{mcitethebibliography}


\providecommand{\latin}[1]{#1}
\makeatletter
\providecommand{\doi}
  {\begingroup\let\do\@makeother\dospecials
  \catcode`\{=1 \catcode`\}=2 \doi@aux}
\providecommand{\doi@aux}[1]{\endgroup\texttt{#1}}
\makeatother
\providecommand*\mcitethebibliography{\thebibliography}
\csname @ifundefined\endcsname{endmcitethebibliography}
  {\let\endmcitethebibliography\endthebibliography}{}
\begin{mcitethebibliography}{0}
\providecommand*\natexlab[1]{#1}
\providecommand*\mciteSetBstSublistMode[1]{}
\providecommand*\mciteSetBstMaxWidthForm[2]{}
\providecommand*\mciteBstWouldAddEndPuncttrue
  {\def\EndOfBibitem{\unskip.}}
\providecommand*\mciteBstWouldAddEndPunctfalse
  {\let\EndOfBibitem\relax}
\providecommand*\mciteSetBstMidEndSepPunct[3]{}
\providecommand*\mciteSetBstSublistLabelBeginEnd[3]{}
\providecommand*\EndOfBibitem{}
\mciteSetBstSublistMode{f}
\mciteSetBstMaxWidthForm{subitem}{(\alph{mcitesubitemcount})}
\mciteSetBstSublistLabelBeginEnd
  {\mcitemaxwidthsubitemform\space}
  {\relax}
  {\relax}

\end{mcitethebibliography}
\end{document}


\doublespace

\maketitle
\thispagestyle{empty}
\begin{abstract}
\noindent
\end{abstract}
\maketitle

\begin{table}[H]
\centering
\begin{tabular}{c|c|c|c|c}
& \multicolumn{2}{c}{Reactant} & \multicolumn{2}{c}{Product} \\
\hline
Bond Harmonic & $K_\mathrm{b}$ & $d_\mathrm{eq}$ & $K_\mathrm{b}$ & $d_\mathrm{eq}$ \\
\hline
N -- H$_\mathrm{NH_{3}}$ & 437.150 & 1.016 & 490.522 &1.039 \\
C -- H$_\mathrm{Me}$ & 375.145 & 1.100 & 369.812 &1.151 \\
\hline
Bond Morse & $D_\mathrm{e}$ & $d_\mathrm{eq}$ & $\beta$ \\
\hline
C -- Cl & 145.720 & 1.785 & 0.582\\
C -- N & 47.9544 & 1.520 & 1.682 \\
\hline
& \multicolumn{2}{c}{Reactant} & \multicolumn{2}{c}{Product}  \\
\hline
Angle & $K_{\Theta}$ & $\Theta_\mathrm{eq}$ & $K_\mathrm{\Theta}$ & $\Theta_\mathrm{eq}$ \\
\hline
H$_\mathrm{NH_{3}}$ -- N -  H$_\mathrm{NH_{3}}$ & 40.594 & 107.500 & 44.274  & 108.806 \\
H$_\mathrm{Me}$ -- C -- H$_\mathrm{Me}$ & 37.170 & 110.010 & 21.686 & 114.437\\
H$_\mathrm{Me}$ -- C -- Cl & 45.335 & 112.040 & X &X \\
H$_\mathrm{NH_{3}}$ -- N -- C & X & X & 57.420 & 107.189 \\
H$_\mathrm{NH_{3}}$ -- C -- H$_\mathrm{Me}$ & X & X & 25.311 & 105.421  \\
\hline
GVDW &  $r$ & $\epsilon$ & $n$ & $m$ \\
\hline
Reactant (N--C) & 1.209 & 0.115 & 6.232 & 12.323 \\
Product (C--Cl)& 2.500 & 0.300 & 5.518 & 12.196 \\
\hline 
\end{tabular}
\caption{The harmonic bond, Morse bond, angle and generalised van der
  Waals (GVDW) parameters for for NH$_{3}$+MeCl. $K_\mathrm{b}$ in
  kcal/mol/\AA\/$^2$, $d_\mathrm{eq}$ in \AA\/, $D_\mathrm{e}$ in
  kcal/mol, $d_\mathrm{eq}$ in \AA\/, $\beta$ in \AA\/$^{-1}$,
  $K_\mathrm{\theta}$ in kcal/mol/radian$^2$, $r$ in \AA\/ and
  $\epsilon$ in kcal/mol.  }
\label{sitab:ff1}
\end{table}

\begin{table}[H]
\centering
\begin{tabular}{c|c|c|c|c}
& \multicolumn{2}{c}{Reactant} & \multicolumn{2}{c}{Product} \\
\hline
Bond Harmonic & $K_\mathrm{b}$ & $d_\mathrm{eq}$ & $K_\mathrm{b}$ & $d_\mathrm{eq}$ \\
\hline
C$_\mathrm{Pyr}$ -- C$_\mathrm{Pyr}$ & 389.240 & 1.369 & 420.800 &1.393 \\
C$_\mathrm{Pyr}$ -- N$_\mathrm{Pyr}$ & 446.700 & 1.319 & 464.880 &1.345 \\
C$_\mathrm{Pyr}$ -- H$_\mathrm{Pyr}$ & 352.560 & 1.083 & 246.970 &1.078 \\
C$_\mathrm{Me}$ -- H$_\mathrm{Me}$ & 293.370 & 1.083 & 226.040 &1.074 \\
\hline
Bond Morse & $D_\mathrm{e}$ & $d_\mathrm{eq}$ & $\beta$ \\
\hline
C -- Br & 87.908 & 1.972 & 0.957\\
C -- N & 286.170 & 1.581 & 0.908 \\
\hline
& \multicolumn{2}{c}{Reactant} & \multicolumn{2}{c}{Product}  \\
\hline
Angle & $K_{\Theta}$ & $\Theta_\mathrm{eq}$ & $K_\mathrm{\Theta}$ & $\Theta_\mathrm{eq}$ \\
\hline
C$_\mathrm{Pyr}$ -- C$_\mathrm{Pyr}$ --  H$_\mathrm{Pyr}$ & 29.657 & 122.100 & 27.082  & 120.800 \\
C$_\mathrm{Pyr}$ -- C$_\mathrm{Pyr}$ --  N$_\mathrm{Pyr}$ & 39.676 & 126.390 & 37.792  & 121.650 \\
N$_\mathrm{Pyr}$ -- C$_\mathrm{Pyr}$ --  H$_\mathrm{Pyr}$ & 43.001 & 118.010 & 52.771  & 115.520 \\
C$_\mathrm{Pyr}$ -- N$_\mathrm{Pyr}$ --  C$_\mathrm{Pyr}$ & 81.961 & 112.800 & 96.761  & 119.040 \\
H$_\mathrm{Me}$ -- C$_\mathrm{Me}$ --  H$_\mathrm{Me}$ & 30.810 & 104.220 & 32.760  & 115.190 \\
H$_\mathrm{Me}$ -- C$_\mathrm{Me}$ --  Br & 40.252 & 102.410 & X  & X \\
N$_\mathrm{Pyr}$ -- C$_\mathrm{Me}$ --  H$_\mathrm{Me}$ & X & X & 48.194  & 104.190 \\
C$_\mathrm{Pyr}$ -- N$_\mathrm{Pyr}$ --  H$_\mathrm{Me}$ & X & X & 40.800  & 121.640 \\
\hline
GVDW &  $r$ & $\epsilon$ & $n$ & $m$ \\
\hline
Reactant (N--C)  & 1.296 & 0.751 &8.096 & 16.710 \\
Product (C--Br)& 0.491 & 0.927 & 7.773 & 11.510 \\
\hline 
\end{tabular}
\caption{The harmonic bond, Morse bond, angle and generalised van der
  Walls (GVDW) parameters for for Pyr+MeBr. $K_\mathrm{b}$ in
  kcal/mol/\AA\/$^2$, $d_\mathrm{eq}$ in \AA\/, $D_\mathrm{e}$ in
  kcal/mol, $d_\mathrm{eq}$ in \AA\/, $\beta$ in \AA\/$^{-1}$,
  $K_\mathrm{\theta}$ in kcal/mol/radian$^2$, $r$ in \AA\/ and
  $\epsilon$ in kcal/mol.}
\label{sitab:ff2}
\end{table}

\begin{table}[H]
\centering
\begin{tabular}{c|c|c|c}
$k$ & $V_\mathrm{ij,k}^{0}$ & $\sigma_\mathrm{ij,k}$ & $\alpha_\mathrm{ij,k0}$ \\
\hline
1&0.48578 & 22.15304 & -17.47731\\
2&40.88660 & 12.45761 & -1.34837 \\
3&-36.67239 & 11.02057 & -1.014567 \\
\end{tabular}
\caption{GAPO parameters for NH$_{3}$+MeCl: $i$ labels the reactant,
  $j$ labels the product, $V_\mathrm{ij,k}^{0}$ is the center of each
  of the $k=3$ Gaussian functions (in kcal/mol),
  $\sigma_\mathrm{ij,k}$ is the width of each Gaussian (in kcal/mol)
  and $\alpha_\mathrm{ij,k0}$ is the polynomial coefficient. In the
  present case polynomial order ``0'' was sufficient, hence
  $\alpha_\mathrm{ij,k0}$.}
\label{sitab:gapo1}
\end{table}

\begin{table}[H]
\centering
\begin{tabular}{c|c|c|c}
$k$ & $V_\mathrm{ij,k}^{0}$ & $\sigma_\mathrm{ij,k}$ & $\alpha_\mathrm{ij,k0}$ \\
\hline
1&-44.46610 & 12.24875 & -0.747557\\
2&1.17231 & 28.18915 & -22.93224 \\
3&55.29621 & 23.35329 & -3.89254 \\
\end{tabular}
\caption{GAPO parameters for Pyr+MeBr: $i$ labels the reactant, $j$
  labels the product, $V_\mathrm{ij,k}^{0}$ is the center of the Gaussian
  function (in kcal/mol), and $\sigma_{ij,k}$ the width of the
  Gaussian (in kcal/mol). $\alpha_{ij,k0}$ is the polynomial
  coefficient in kcal/mol. In the present case polynomial order
  ``0'' was sufficient, hence $\alpha_\mathrm{ij,k0}$.}
\label{sitab:gapo2}
\end{table}

\begin{figure}[H]
\begin{center}
\includegraphics[width=0.9\linewidth]{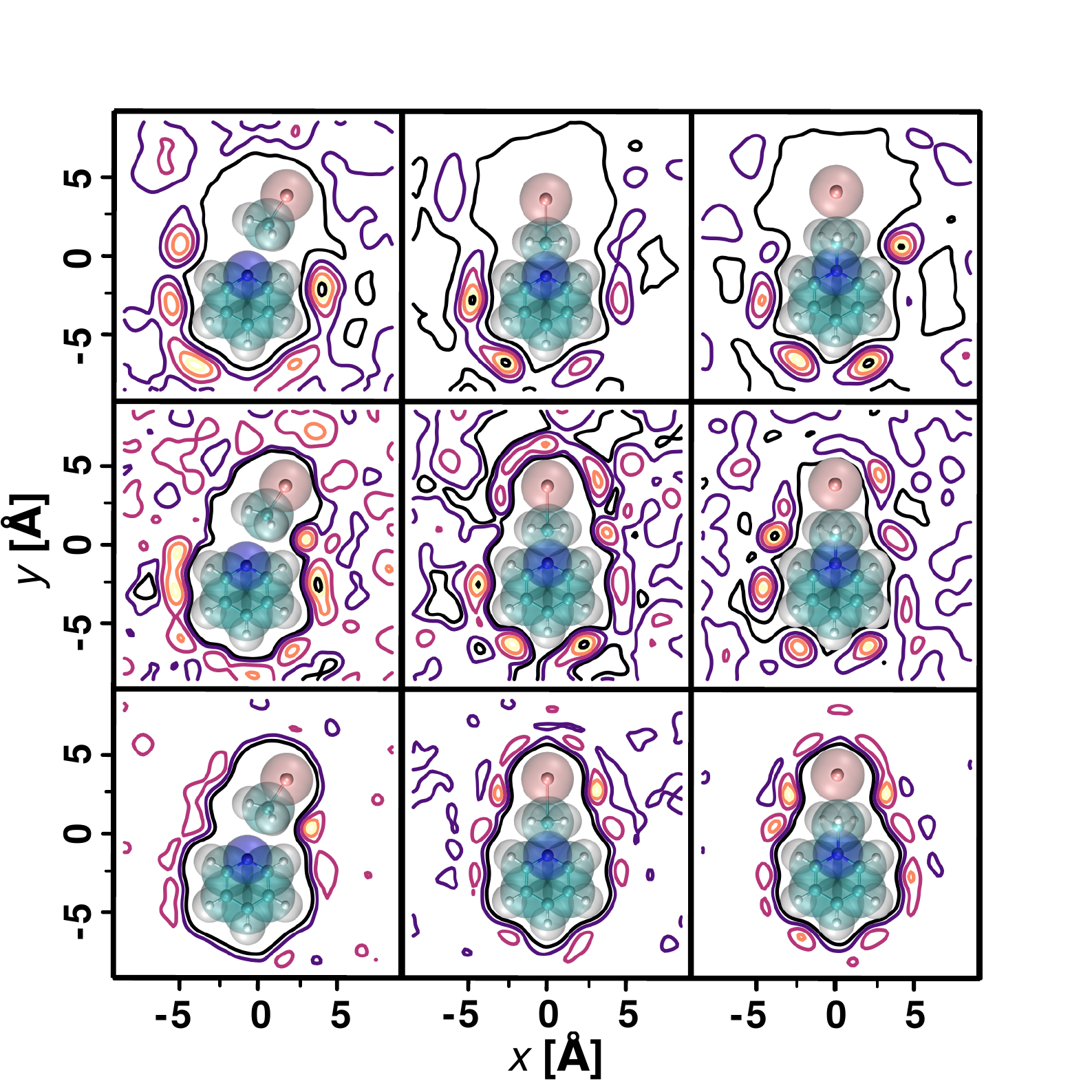}
\caption{2-dimensional solvent distributions for Pyr+MeBr from 2 ns
  simulations in polar solvents: acetonitrile, methanol and water
  (from top to bottom) around the reactant, TS, and product state
  structures of the solute (from left to right) projected onto the
  $xy-$plane containing the chloride, carbon and nitrogen atoms. Units
  in \AA\/. In the simulations and the figure the solute is in its
  optimized structure for the reactant, TS, and product state,
  respectively, at the MP2/6-311++G(2d,2p) level of theory. The color
  code for atoms is H (white), C (cyan), N (blue) and Br (pink).}
\label{sifig:mebr_polar}
\end{center}
\end{figure}

\begin{figure}[H]
\begin{center}
\includegraphics[width=\linewidth]{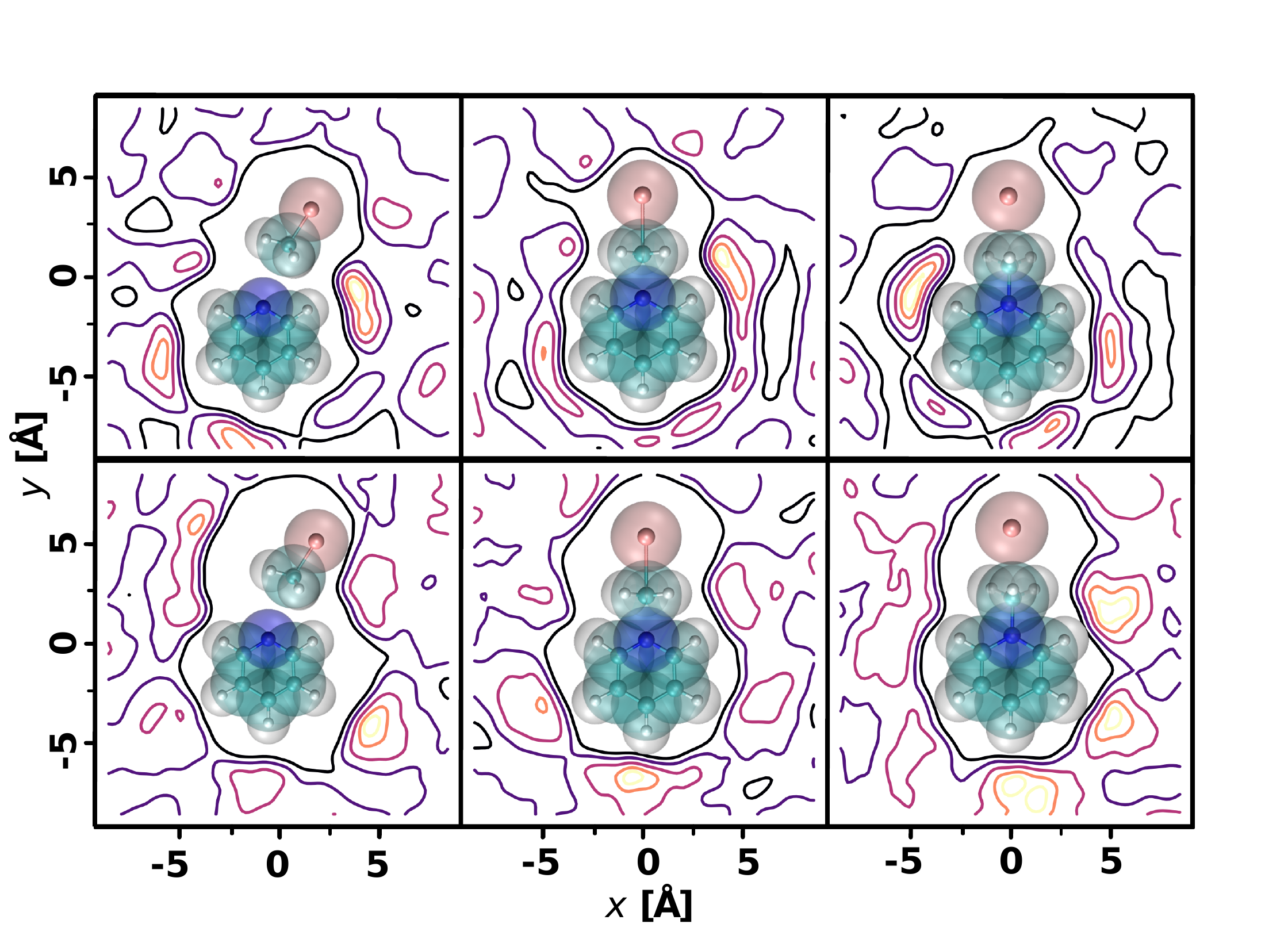}
\caption{2-dimensional solvent distributions for Pyr+MeBr in apolar
  solvents: benzene (top) and hexane (bottom) around the reactant, TS,
  and product state structures of the solute (from left to right)
  projected onto the $xy-$plane containing the chloride, carbon an
  nitrogen atoms. Units in \AA\/. In the simulations and the figure
  the solute is in its optimized structure for the reactant, TS, and
  product state, respectively, at the MP2/6-311++G(2d,2p) level of
  theory. The color code for atoms is H (white), C (cyan), N (blue)
  and Br (pink).}
\label{sifig:mebr_apolar}
\end{center}                
\end{figure}

\begin{figure}[H]
\begin{center}
\includegraphics[width=0.8\linewidth]{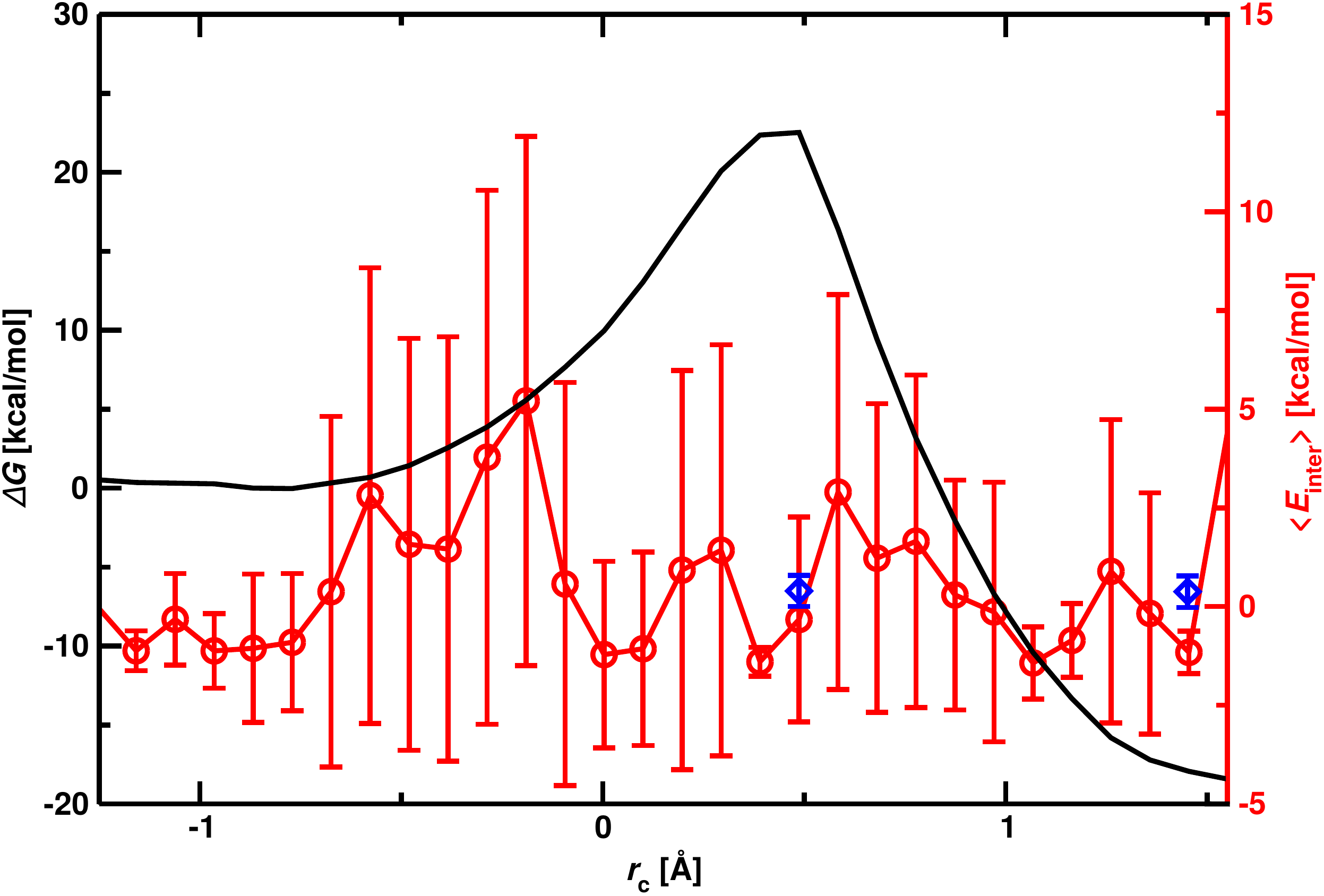}
\caption{Potentials of mean force for Pyr+MeBr in methanol (black
  line) and the average solvent-solvent interaction for molecules
  5\AA\/ around the solute in kcal/mol/molecule (red line). The
  circles are the mean for a given umbrella together with the standard
  deviation from the mean (fluctuation bar). The blue diamonds are
  average solvent-solvent interaction energies for molecules 5\AA\/
  around the solute in kcal/mol/molecule from 2 ns $NVT$ simulations
  in reactant, TS and product states, respectively, together with the
  standard deviation (fluctuation bar).}
\label{sifig:small_box_MeBr}
\end{center}
\end{figure}

\begin{figure}[H]
\begin{center}
\includegraphics[width=\linewidth]{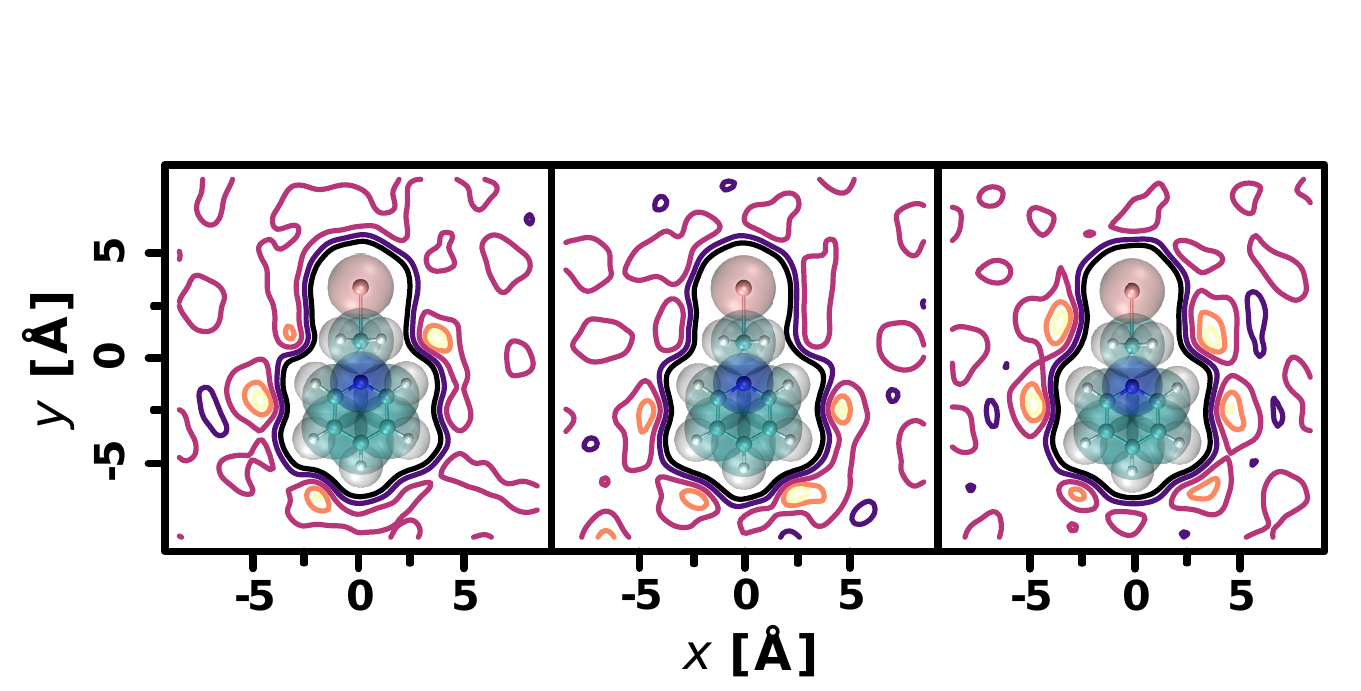}
\caption{2-dimensional solvent distribution for Pyr+MeBr in methanol
  for $r_{c} = [-0.3, -0.2, -0.1]$ projected onto the $xy-$plane,
  respectively. The color code for atoms is H (white), C (cyan), N (blue)
  and Br (pink).}
\label{sifig:mebr_umb_solv}
\end{center}                
\end{figure}